%% file: main.tex
\documentclass[journal]{config/new-aiaa}
\input{config/config}
\input{config/front_matter}

\begin{document}

\maketitle

\begin{abstract}
In this paper, we present Iterative Classification of Graph-Set-Based Design (IC-GSBD), a framework utilizing graph-based techniques with geometric deep learning (GDL) integrated within a set-based design (SBD) approach for the classification and down-selection complex engineering systems represented by graphs.
We demonstrate this approach on aircraft thermal management systems (TMSs) utilizing previous datasets created using an enumeration or brute-force graph generation procedure to represent novel aircraft TMSs as graphs.
However, as with many enumerative approaches, combinatorial explosion limits its efficacy in many real-world problems, particularly when simulations and optimization must be performed on the many (automatically-generated) physics models. 
Therefore, the approach uses the directed graphs representing aircraft TMSs and GDL to predict on a subset of the graph-based dataset through graph classification.
This paper's findings demonstrate that incorporating additional graph-based features using principle component analysis (PCA) enhances GDL model performance, achieving an accuracy of 98\% for determining a graph's compilability and simulatability while using only 5\% of the data for training. 
By applying iterative classification methods, we also successfully segmented the total set of graphs into more specific groups with an average inclusion of 75.5 of the top 100 highest-performing graphs, achieved by training on 40\% of the data.
\end{abstract}

\input{nomenclature}


\section{Introduction}
\input{input/introduction}
\label{sec:intro}

\section{Background}
\label{background}
\input{input/background/intro}

\section{Methodology}
\label{methods}
\input{input/methodology/intro}

\section{Case Study Scenarios and Results for Aircraft Thermal Management System Design}
\label{results}
\input{input/results/intro}

\section{Conclusion}
\label{conclusion}
\input{input/conclusion}

\bibliography{references_final}

\end{document}

%% file: config/config.tex
\usepackage[utf8]{inputenc}
\usepackage{textcomp}
\usepackage{graphicx}
\usepackage{amsmath}
\usepackage[version=4]{mhchem}
\usepackage{siunitx}
\usepackage{longtable,tabularx}
\usepackage{bm}
\usepackage{nicematrix}
\usepackage{soul}
\usepackage{xcolor}
\usepackage{booktabs,subcaption,amsfonts,dcolumn}
\usepackage{bm}
\usepackage{nicematrix}
\usepackage[normalem]{ulem}
\usepackage{ltxcmds}
\usepackage{mathtools}
\usepackage{multirow} 
\usepackage{enumitem}
\usepackage{ifthen}
\usepackage{etoolbox}
\usepackage{cancel}
\usepackage{empheq}
\usepackage{tikz}
\usepackage[framemethod=TikZ]{mdframed}

\usepackage{xpatch}
\usepackage{dblfloatfix}
\usepackage{float}
\usepackage{xcolor}
\definecolor{lightblue}{RGB}{173, 216, 230}
\definecolor{lightred}{RGB}{255, 182, 193}
\usepackage{colortbl}
\usepackage{accents}
\usepackage{adjustbox}
\usepackage{amsmath}
\usepackage{bm}
\usepackage{nicematrix}
\usepackage{varwidth}
\usepackage{bbold}
\usepackage{algpseudocode}
\usepackage{algorithm}
\usetikzlibrary{shapes.geometric, arrows, positioning, arrows.meta, backgrounds, fit}

\tikzstyle{startstop} = [rectangle, rounded corners, minimum width=3cm, minimum height=1cm, text centered, draw=black, fill=red!30, font=\small]
\tikzstyle{process} = [rectangle, minimum width=3cm, minimum height=1cm, text centered, draw=black, fill=orange!30, text width=0.3\columnwidth, font=\small]
\tikzstyle{decision} = [diamond, aspect=2, minimum width=3cm, minimum height=1cm, text centered, draw=black, fill=green!30, text width=3cm, font=\small]
\tikzstyle{arrow} = [thick,->,>=stealth]
\tikzstyle{dygide} = [rectangle, dashed, minimum width=3cm, minimum height=1cm, text centered, draw=black, fill=blue!10, text width=0.3\columnwidth, font=\small]
\tikzstyle{automate} = [rectangle, dashed, minimum width=3cm, minimum height=1cm, text centered, draw=black, fill=red!10, text width=0.3\columnwidth, font=\small]

\newcommand{\eref}[1]{Eq.~\eqref{#1}}        
\newcommand{\fref}[1]{Fig.~\ref{#1}}   
\newcommand{\cref}[1]{Chap.~\ref{#1}}  
\newcommand{\sref}[1]{Sec.~\ref{#1}}  
\newcommand{\tref}[1]{Table~\ref{#1}}    
\newcommand{\rref}[1]{Ref.~\cite{#1}}
\newcommand{\alref}[1]{Algorithm~\ref{#1}}

\newcommand{\xz}[1]{\textcolor{lightgray}{#1}}
\newcommand{\xy}[1]{\scriptstyle \textsf{#1}}

\definecolor{poscolor}{HTML}{1e88e5}
\definecolor{negcolor}{HTML}{e53935}
\definecolor{offcolor}{HTML}{A93C93}

\newcommand*{\rmFullStop}{\rmifnextchar{.}{}{}}

\makeatletter
\newcommand{\rmifnextchar}[3]{%
  \begingroup
  \ltx@LocToksA{\endgroup#2}%
  \ltx@LocToksB{\endgroup#3}%
  \ltx@ifnextchar{#1}{%
    \def\next{\the\ltx@LocToksA}%
    \afterassignment\next
    \let\scratch= %
  }{%
    \the\ltx@LocToksB
  }%
}
\makeatother

\hypersetup{
    colorlinks=true,
    allcolors=blue,
    pdftitle={Geometric Deep Learning Towards the Iterative Classification of Graph-Based Aircraft Thermal Management Systems},
    pdfauthor={Anthony Sirico Jr., Daniel R. Herber},
}

\usepackage{subcaption}

\newcommand{\doi}[1]{{doi:~\href{http://doi.org/#1}{#1}}\rmFullStop}

%% file: config/front_matter.tex
\title{Iterative Classification of Graph-Set-Based Designs (IC-GSBD) for the Down-Selection of Aircraft Thermal Management Systems}

\author{Anthony Sirico Jr.\footnote{Ph.D., Systems Engineering, 200 W. Lake Street, 6029 Campus Delivery, Fort Collins, CO 80523} and Daniel R. Herber\footnote{Assistant Professor, Systems Engineering, 200 W. Lake Street, 6029 Campus Delivery, Fort Collins, CO 80523}}
\affil{Colorado State University, Fort Collins, CO 80523}

%% file: nomenclature.tex
\section*{Nomenclature}

{\renewcommand\arraystretch{1.0}
\noindent\begin{longtable*}{@{}l @{\quad=\quad} l@{}}
$\mathbf{A}$ & An adjacency matrix\\
$a_{i,j}$ & A values from an adjacency matrix at row $i$, column $j$\\
ACM & Air Cycle Machine\\
ACP & Air Compressor\\
BSI & Bleed Source In\\
BSO & Bleed Source Out\\
$C$ & Harmonic Centrality\\
$c_B$ & Betweenness Centrality\\
$C_{total}$ & The total value of the heat exchangers\\
CNN & Convolutional Neural Network\\
DGCNN & Deep Graph Convolutional Neural Network\\
DyGIDE & Dynamic Geo-Iterative Design Exploration\\
$E$ & A set of graph edges\\
$F_P$ & False Positive\\
$F_N$ & False Negative\\
$\mathcal{G}$ & A set of graphs\\
$\mathcal{G}_{all}$ & An set of graphs that will be split into ``known'' and ``unknown'' sets\\
$\mathcal{G}_{known}$ & The set of graphs that was analyzed\\
$|\mathcal{G}_{known}|_{min}$ & The minimum value for the size of $\mathcal{G}_{known} $ at iteration $n$\\
$\mathcal{G}_{training}$ & The set of graphs earmarked for training the model\\
$\mathcal{G}_{unknown}$ & The set of graphs that was not analyzed and will be predicted on\\
$\mathcal{G}_{validation}$ & The set of graphs earmarked to test the model during training\\
\(G\) & A single graph\\
GDL & Geometric Deep Learning\\
GNN & Graph Neural Network\\
HEX & Heat Exchanger\\
$\bar{J}$ & Median performance value\\
\(J(G_i)\) & Performance Value of a graph\\
$k$ & Iteration number\\
$\lambda$ & An eigenvalue\\
$\mathbf{L}$ & A graph label array\\
$l_c$ & A single graph label from the set of graphs labels $\mathbf{L}$\\
$m$ & A Geometric Deep Learning model\\
$\Dot{m}$ & Mass flow rate\\
MCC & Matthews Correlation Coefficient\\
$n$ & The number of iterations to be completed during iterative classification\\
$N_{all}$ & The size of the set $\mathcal{G}_{all}$\\
$N_{known}$ & The number of graphs in the ``known'' dataset\\
$N_{unknown}$ & The number of graphs in the ``unknown'' dataset\\
$\varphi$ & A one-to-one mapping between two sets of of graph vertices $V_1$ and $V_2$\\
PCA & Principle Component Analysis\\
$\rho$ & Spectral Radius\\
$\mathbf{r}$ & The output vector of the Global Mean Pooling layer\\
RSI & Ram-air Source In\\
RSO & Ram-air Source Out\\
SBD & Set-Based Design\\
SVD & Singular Value Decomposition\\
$T_P$ & True Positive\\
$T_N$ & True Negative\\
TMS & Thermal Management System\\
TUR & Turbine\\
$V$ & A set of graph vertices\\
$\mathbf{W}$ & The weights matrix\\
$w$ & Calculated Weight\\
$\mathbf{X}$ & A features matrix for a graph $G$\\
\end{longtable*}}

%% file: input/introduction.tex
With hybrid and electric aircraft on the horizon, aircraft flying at higher Mach numbers, and the ever-increasing required aircraft capabilities, the need for more efficient cooling methods increases substantially. 
The modernization of aircraft has also led to more efficient engines that burn less fuel per pound of thrust, reducing the cooling capacity available during operation.
Furthermore, the use of composite materials that have a reduced ability to dissipate heat \cite{heerden2022, huang2004, coutinho2023}.
This evolving landscape necessitates the call for a revolution in the overall design process to arrive at novel solutions.
With this high demand for improved aircraft thermal management systems (TMSs), researchers have turned to different ways to generate potential design options that meet specific requirements, including levels of automation \cite{herber2020, buettner2021}.

Recent advancements in engineering design methodologies have increasingly emphasized the need for more comprehensive and flexible approaches to navigate the complexities of system design and optimization.
The Set-Based Design (SBD) philosophy of broad exploration of alternatives followed by gradual refinement is particularly resonant in complex systems design, where the interaction between various components and constraints significantly influences the final design outcome.
To encapsulate those interactions and constraints of the components, we represent these systems as mathematical graphs.

Mathematical graphs are highly effective in representing various systems and decisions because they encapsulate discrete compositional and relational information. 
For decades, numerous studies have employed different graph representations to tackle specific problems \cite{herber2017, b6, b37, b95, b96, b97, b102, b105, herber2020, arney2014}. 
Researchers have leveraged graph enumeration for over a century to comprehend engineering design challenges and support decision-making \cite{herber2017_2, herber2017_3, b6, b37, b38}. 
More recently, many engineering design issues, including the construction of the "system architecture," have expanded in scope and complexity to the extent that traditional discrete and continuous representations fall short of adequately depicting the system \cite{b6}.
Graph enumeration generates a comprehensive and ordered list of possible graphs for a given structure \cite{herber2017_2, b30, b31}. 
But, depending on the specific problem and the selected representation, this method can yield many potential solutions. 
Consequently, computational costs have increased exponentially, complicating the decision-making process for designers or system architects.

Therefore, to enhance the engineering design process by combining the SBD philosophy, the need for automation, and graph-based methods, the framework Graph-Set-Based Design (GSBD) was introduced in \rref{Sirico2024c} as well as Iterative Classification (IC) in response to the need for a method for determining more promising and effective solutions.
When combined with GSBD, becoming IC-GSBD, this methodological framework offers a novel approach to system design and optimization.

As a major final component, IC-GSBD utilizes Geometric Deep Learning (GDL), a deep learning method that generalizes neural networks to both Euclidean and non-Euclidean domains, to include graphs, manifolds, meshes, and string representations \cite{b19}, utilizing a Graph Neural Network (GNN).
In essence, GDL encompasses approaches that incorporate information on the input variables' structure space and symmetry properties and leverage it to improve the quality of the data captured by the model.
The appropriate construction and use of GDL can result in more efficient ways of predicting the value or performance of an architecture represented as a graph.

Amidst this, IC-GSBD's strategic integration of GDL can significantly augment the SBD process by enhancing the efficiency and depth of the exploratory phase and addressing the computational intensiveness of the traditional SBD process, making IC-GSBD an indispensable advancement, streamlining the design process by leveraging GDL's robust analytical capabilities\cite{singer2009, parnell2019}.

The essence of IC-GSBD lies in the synergy between GDL and SBD. 
While SBD allows for the exploration of design sets underpinned by varying parameters and constraints, IC-GSBD enriches this exploration by quickly assessing each set's viability and performance potential. 
This unique capability ensures a more informed selection process by dynamically adapting to new insights and data, effectively reducing the iterative cycles typically required in SBD.
This capability is crucial in early design stages, where the ability to rapidly gauge the impact of design decisions can significantly streamline the design process, reduce computational overhead, and guide the focus towards the most promising design avenues.

Moreover, IC-GSBD embodies the SBD philosophy by enabling a systematic and iterative exploration of the design space, yet it transcends traditional limitations by leveraging GDL's analytical prowess.
This leap forward is achieved by utilizing advanced machine learning to interpret and analyze the intricate relationships within design parameters. 
Thus, the depth of analysis is significantly elevated beyond what is conventionally possible with SBD alone.
In doing so, IC-GSBD manages complexity more effectively and enhances the efficiency and depth of the design exploration process. 
IC-GSBD stands as a testament to the potential of combining advanced machine learning techniques with established design methodologies to navigate the challenges inherent in designing complex engineering systems like aircraft TMS.

Graph-based methods are a promising way of representing architecture candidates \cite{herber2017, herber2020, selva2016, arney2014}, and the use of GDL is the basis for the down-selecting process of these graph-based engineering designs using iterative classification \cite{sirico2023}.
Previously in \rref{sirico2023}, this method significantly narrowed down 43,000+ potential graphs to 87 of the top 100 top-performing designs with 1/4 the computational cost.
The primary motivation behind this investigation is similar, i.e., to introduce automation into this process of generating candidate architectures while reducing the computational cost.

By extending the capabilities of SBD through a GDL-based framework, IC-GSBD offers a more dynamic, informed, and efficient approach to designing and selecting aircraft TMS. 
This paradigm shift redefines the boundaries of complex systems design and demonstrates a scalable model for applying advanced computational techniques to traditional design methodologies.

The remainder of this paper is as follows. 
First, in \sref{background}, we discuss SBD, the graph-based TMS models, and the necessary background information to understand both IC-GSBD and GDL.
Next, in \sref{methods}, we discuss how IC-GSBD was implemented and the various metrics used to determine a well-performing machine learning model.
In \sref{case}, we apply IC-GSBD to an engineering design case study of selecting a physics-based TMS graph for unmanned aerial vehicles (UAV) and discussing the outcomes.
Finally, the concluding remarks are expressed in \sref{conclusion}.

%% file: input/background/intro.tex
This section aims to deepen the understanding of the underlying motivation and approaches of the proposed methodology.
In our discussion, we will delve into the basics of Set-Based Design (SBD) and how it can be further strengthened by integrating the DyGIDE framework. 
We will then provide an overview of Geometric Deep Learning (GDL), covering important concepts such as graph theory and Graph Neural Networks (GNNs). 
This will help us understand the fundamentals of GDL and its potential applications.

A significant hurdle in using graphs for engineering design is the occasionally prohibitive computational expense of evaluating $J(G_i)$. 
This evaluation can be more costly than generating the graph $G_i$. 
The origins of these expenses are varied, encompassing high-fidelity simulations, optimization processes, evaluations involving human judgment, and physical experimentation. Our work concentrates on scenarios where these costs become a limiting factor.

Based on the classification problem of \rref{herber2020}, we define the following three types of graph-centric design problems:
\begin{enumerate}[nolistsep,label=$\bullet$]
   
\item \textit{Type 0} --- All desired graphs can be generated, and so can their performance metric $J(G_i)$ within time $T$

\item \textit{Type 1} --- All desired graphs can be generated, but only some of the performance metrics $J(G_i)$ can be evaluated within time $T$; the performance assessment is too expensive

\item \textit{Type 2} --- All desired graphs cannot be generated within time $T$
    
\end{enumerate}

\noindent where $ T$ represents the time allocated to complete the graph design study. 
We will focus on the Type 1 problem (using data from a large Type 0 study).

\subsection{Set-Based Design (SBD)}
\label{sec:sbd}
\input{input/background/sbd}

\subsection{Graph Theory}
\label{graphtheory}
\input{input/background/graph_theory}

\subsection{Representation, Enumeration, and Analysis of Thermal Management Systems as Graphs}
\label{enum}
\input{input/background/tms_models_enumeration}


\subsection{Geometric Deep Learning}
\label{gdl}
\input{input/background/gdl}

\subsection{Graph Neural Networks}
\label{gnn}
\input{input/background/gnn}

%% file: input/background/sbd.tex
Set-Based Design (SBD) is a transformative design methodology that underscores the importance of exploring a wide array of potential solutions at the outset of the design process. 
Unlike traditional Point-Based Design (PBD), which focuses on iteratively refining a singular design solution within a predefined set of constraints, SBD facilitates a comprehensive exploration of the design space. 
This approach enables designers to concurrently consider multiple alternatives and their interrelations, fostering innovation and leading to more optimized solutions \cite{singer2009, parnell2019}.

The utilization of an SBD framework presents several advantages:

\begin{itemize}
\item \textbf{Concurrent Consideration of Alternatives:} It encourages the exploration of diverse design solutions simultaneously.
\item \textbf{Early Identification of Constraints and Requirements:} Evaluating a wide range of solutions helps to identify constraints and requirements early on in the design process.
\item \textbf{Enhanced Decision-Making:} SBD allows for an in-depth assessment of trade-offs between different design alternatives, facilitating more informed decision-making.
\item \textbf{Reduction of Design Iterations:} It mitigates the risk of prematurely converging on a single design solution, potentially reducing the number of iterations needed to reach an optimized design.
\end{itemize}

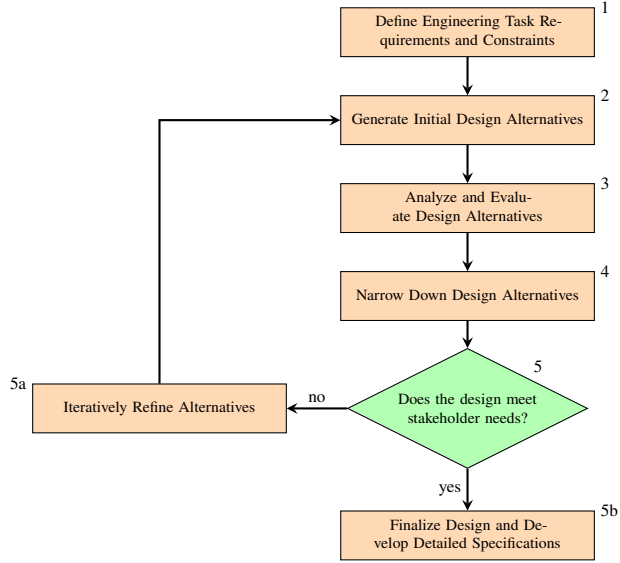
\begin{figure}[t]
    \centering
    \begin{tikzpicture}[node distance=1.8cm, scale=0.65, every node/.style={transform shape}]
        \node (start) [process] {Define Engineering Task Requirements and Constraints};
        \node (generate) [process, below of=start] {Generate Initial Design Alternatives};
        \node (analyze) [process, below of=generate] {Analyze and Evaluate Design Alternatives};
        \node (narrow) [process, below of=analyze] {Narrow Down Design Alternatives};
        \node (decision) [decision, below of=narrow, yshift=-0.5cm] {Does the design meet stakeholder needs?};
        \node (refine) [process, left of=decision, xshift=-4.5cm] {Iteratively Refine Alternatives};
        \node (final) [process, below of=decision, yshift=-0.8cm] {Finalize Design and Develop Detailed Specifications};

        \node[right] at (start.north east) {1};
        \node[right] at (generate.north east) {2};
        \node[right] at (analyze.north east) {3};
        \node[right] at (narrow.north east) {4};
        \node[above right] at (decision.north east) {5};
        \node[left] at (refine.north west) {5a};
        \node[right] at (final.north east) {5b};   
        
        \draw [arrow] (start) -- (generate);
        \draw [arrow] (generate) -- (analyze);
        \draw [arrow] (analyze) -- (narrow);
        \draw [arrow] (narrow) -- (decision);
        \draw [arrow] (decision) -- node[anchor=east] {yes} (final);
        \draw [arrow] (decision) -- node[anchor=south] {no} (refine);
        \draw [arrow] (refine) |- (generate);
    \end{tikzpicture}
    
    \caption{The workflow for Set-Based Design.}
    \label{fig:sbdflow}
\end{figure}


As highlighted in \rref{Ward1995, Sobek1999, Kennedy2009}, the Set-Based Design (SBD) process encompasses a structured approach to engineering design as shown in \fref{fig:sbdflow}.
Initially, it involves defining the requirements and constraints that shape the exploration of potential design solutions (Step 1), ensuring all considered alternatives align with the project's overarching goals.
Following this, a wide array of design alternatives is developed (Step 2), deliberately avoiding early commitment to any single solution.
Subsequently, parallel analysis and evaluation of these alternatives are conducted against the set requirements and constraints (Step 3), a step underscored for its importance in iterative refinement and decision-making.
This analysis helps progressively narrow down the options based on their feasibility and alignment with project goals (Step 4). 
The narrowing process is inherently iterative, requiring repeated analysis and evaluation to continually refine the set of alternatives until a final design is selected (Step 5). 
This chosen design then undergoes the development of detailed specifications (Step 5b), marking the transition from a conceptual stage to detailed engineering and production planning.
If the design does not meet stakeholder needs, alternatives are iteratively refined (Step 5a).

%% file: input/background/graph_theory.tex
In the previous section, we talked about representing architectures as graphs. 
In this section, we will learn about the fundamentals of graph theory to understand how systems, especially thermal management systems, can be represented as graphs. We will start by defining some key terms.

\vspace{\baselineskip}
\noindent \textbf{Definition 1 (Graph)}. A \textit{graph} $G$ is a pair of sets $(V,E)$ where $E \subseteq [V]^2$ (i.e., $E$ is a two-element subset of $V$). $E$ represents the edges of the graph, while $V$ is the set of its vertices or nodes \cite{diestel2017}.
While there are different ways to represent a graph mathematically, here we focus on one known as the \textit{adjacency matrix} $\textbf{A} = (a_{i,j})_{n \times n}$ and is defined as:
\begin{align}
a_{i,j} \coloneqq 
\begin{cases}
1, & \text{ if } (v_i,v_j) \in E \\
0, & \text{ otherwise}
\end{cases}
\label{adjeq}
\end{align}
where $n$ represents the number of nodes in the graph, and the pair $(i,j)$ represents the row and column node indices in the matrix.
\begin{figure}
    \centering
    \begin{subfigure}[t]{0.42\textwidth}
        \includegraphics[scale=0.85]{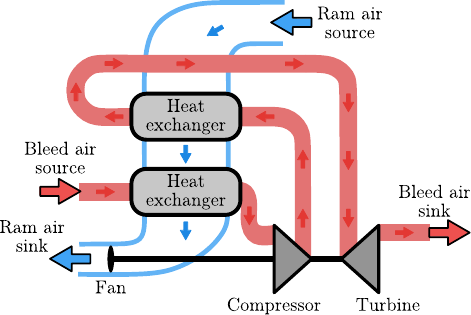}
        \caption{Illustration.}
        \label{acm}
    \end{subfigure}
    \begin{subfigure}[t]{0.42\textwidth}\centering
        \includegraphics[scale=0.85]{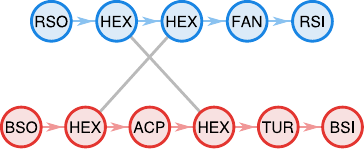}
        \caption{Graph representation.}
        \label{acmgraph}
    \end{subfigure}

    \caption{A common air cycle machine (ACM) architecture represented in two forms.}
    \label{fig:acmandgraph}
\end{figure}

\vspace{\baselineskip}
\noindent \textbf{Definition 2 (Directed Graph).} A directed graph, or digraph, is a pair $(V,E)$ of disjoint sets together with two maps init:$E\to V$ and ter:$E\to V$ assigning to every edge $e$ an initial node init$(e)$ and a terminal node ter$(e)$ \cite{diestel2017}. 
This orientation arises when one node acts as a ``source'' node and is directed towards another.
If we examine \fref{acmgraph}, it becomes apparent that in a directed graph, connections are represented by directed arrows instead of lines, indicating the direction of flow from the source node. 
Furthermore, as shown in \eref{digrapheq} representing the adjacency matrix for \fref{acmgraph}, a value of 1 signifies the existence of a connection from the source node.
{
    \setlength{\arraycolsep}{2pt}
    \renewcommand{\arraystretch}{0.8}
    \begin{equation}
        \scalebox{0.9}{$
        \mathbf{A} = 
        \begin{bNiceArray}{ccccccccccc}[last-row,last-col]
           \xz{0} & 1 & \xz{0} & \xz{0} & \xz{0} & \xz{0} & \xz{0} & \xz{0} & \xz{0} & \xz{0} & \xz{0} & \xy{RSO}\\
           \xz{0} & \xz{0} & 1 & \xz{0} & \xz{0} & \xz{0} & \xz{0} & \xz{0} & \xz{0} & \xz{0} & \xz{0} & \xy{HEX}\\
           \xz{0} & 1 & \xz{0} & 1 & \xz{0} & \xz{0} & 1 & \xz{0} & 1 & \xz{0} & \xz{0} & \xy{HEX}\\
           \xz{0} & \xz{0} & \xz{0} & \xz{0} & 1 & \xz{0} & \xz{0} & \xz{0} & \xz{0} & \xz{0} & \xz{0} & \xy{HEX}\\
           \xz{0} & \xz{0} & \xz{0} & 1 & \xz{0} & \xz{0} & \xz{0} & \xz{0} & \xz{0} & 1 & \xz{0} & \xy{HEX}\\
           \xz{0} & \xz{0} & \xz{0} & 1 & \xz{0} & \xz{0} & \xz{0} & \xz{0} & \xz{0} & \xz{0} & \xz{0} & \xy{BSO}\\
           \xz{0} & \xz{0} & \xz{0} & \xz{0} & \xz{0} & \xz{0} & \xz{0} & 1 & \xz{0} & \xz{0} & \xz{0} & \xy{FAN}\\
           \xz{0} & \xz{0} & \xz{0} & \xz{0} & \xz{0} & \xz{0} & \xz{0} & \xz{0} & \xz{0} & \xz{0} & \xz{0} & \xy{RSI}\\
           \xz{0} & \xz{0} & \xz{0} & \xz{0} & 1 & \xz{0} & \xz{0} & \xz{0} & \xz{0} & \xz{0} & \xz{0} & \xy{ACP}\\
           \xz{0} & \xz{0} & \xz{0} & \xz{0} & \xz{0} & \xz{0} & \xz{0} & \xz{0} & \xz{0} & \xz{0} & 1 & \xy{TUR}\\
           \xz{0} & \xz{0} & \xz{0} & \xz{0} & \xz{0} & \xz{0} & \xz{0} & \xz{0} & \xz{0} & \xz{0} & \xz{0} & \xy{BSI}\\ 
           \xy{RSO} & \xy{HEX} & \xy{HEX} & \xy{HEX} & \xy{HEX} & \xy{BSO} & \xy{FAN} & \xy{RSI} & \xy{ACP} & \xy{TUR} & \xy{BSI} & =\mathbf{X}
        \end{bNiceArray}
        $}
        \label{digrapheq}
    \end{equation}
}

Another important concept is graph isomorphism. Consider two graphs, \( G_1 = (V_1,E_1) \) and \( G_2 = (V_2,E_2) \). 
These graphs are termed isomorphic, expressed as \( G1 \cong G_2 \), if a bijection, or one-to-one mapping, \( \varphi \) exists from \( V_1 \) to \( V_2 \) such that for every pair of vertices \( (v_i,v_j) \) in \( E_1 \), the corresponding pair \( (\varphi(v_i),\varphi(v_j)) \) is in \( E_2 \). 
Furthermore, the concept extends to labeled graph isomorphism. 
Using the same two graphs, if they include an additional feature such as a vertex label \( X \), making the graphs \( G_1 = (V_1,E_1,X_1) \) and \( G_2 = (V_2,E_2,X_2) \), they are considered isomorphic if this vertex label characteristic is maintained under a valid bijection \( \varphi \). 
In a broader context, a feature matrix \( \textbf{X} \in \mathbb{R}^{n \times c} \) can include numerous columns to represent the \( c \) features associated with each vertex.

%% file: input/background/tms_models_enumeration.tex
Graph-based models can represent TMS architectures using directed labeled graphs, depicting physical/functional components as vertices and connections as edges \rref{herber2020}, as illustrated in Fig.~\ref{fig:acmandgraph}.
Here, we will briefly review the methods from \rref{herber2020, buettner2021} that generate and evaluate these graphs that are the focus of the case study.

\subsubsection{Enumerative Methods with Network Structure Constraints}

Enumerative techniques for graph theory involve creating a comprehensive catalog of \textit{all} graphs that meet specific criteria. 
This complete catalog is often referred to as a graph structure space.
The original study in \rref{herber2020} used multiple enumeration methods to ensure that all directed labeled graphs of interest were generated.
With efficient enumeration graph algorithms, the relevant graph structure spaces can be determined (e.g., including hundreds of thousands of unique graphs).


To help improve the usefulness of the generated graphs, \textit{Network Structure constraints} (NSCs) were included in the generation and post-filtering steps.
These include the direct connection NSC, which restricts connections between certain vertex types; the line-connectivity NSC, specifying permissible connections between different vertex types; and the multi-edges NSC, which ensures each edge in the graph is unique. 
Additionally, the intermediate component NSC identifies specific vertex types within a path, the downstream vertex NSC determines vertex relationships in a directed path graph, and the vertex-subgraph NSC manages vertex distribution across different subgraphs. 
Each constraint contributes significantly to the structural integrity and feasibility of the generated graphs \cite{herber2017_2, herber2017_3}.
However, even with these NSCs, there are still valid graphs that are not valid models of the physically realistic TMS architectures, which can be determined through physics-based analysis.

\subsubsection{Analysis of Physics-Based Models}

The graphs from the previous section now serve as data objects for constructing detailed simulation models in Modelica, enabling engineers to assess the performance and realizability of the TMS architecture.
This process entails mapping each vertex in the graph to a specific Modelica component model, transforming the abstract graph representation into a text-based Modelica model file system that can be compiled and simulated to help understand how the TMS will perform under different conditions.

Each TMS component (e.g., turbines, compressors, heat exchangers, etc.) is linked to a physics-based model capturing its behavior based on fundamental physical principles and data.
For instance, a basic air cycle machine (ACM), as shown in \fref{acm}, consists of two heat exchangers (HEX), a compressor (ACP), a turbine (TUR), and a fan (FAN). This system incorporates two primary air sources: ram air (RSO) and bleed air (BSO). 
The RSO utilizes external air, considerably cooler than the air inside the aircraft at high altitudes, to cool various components and engine operations. 
Bleed air, on the other hand, is sourced from the engine's compressor section and plays a vital role in the aircraft's environmental control and anti-ice systems. 
Incoming ram air is passed through the heat exchangers to help cool the engine bleed air as it passes from the compressor to the turbine back to the bleed air sink.
Now observing \fref{acmgraph}, we can see how the basic ACM can be represented as a graph, where the edges show the various air flows.


%% file: input/background/gdl.tex
Graphs fundamentally differ from the more common Euclidean data in most deep-learning applications (e.g., images, text, and speech)\cite{NIPS2012_c399862d, wang2012, deng2013}.
Euclidean data has an underlying grid-like structure.
For example, an image can be translated to an $(x,y,z)$ Cartesian coordinate system, where each pixel is located at an $(x,y)$ coordinate, and $z$ represents the color. 
But, this approach only works for some problems because certain operations require many dimensions.
This ``flat'' representation has limitations, such as being unable to represent hierarchies and other direct relationships.

These issues motivated the study of hyperbolic space for graph representation, or non-Euclidean learning \cite{nickel2017, chamerlain2017}, which could perform those same operations more flexibly.
For example, many real-world data sets are naturally structured as graphs, where the \textit{relationships} between data points or other entities are integral.
Euclidean-based methods struggle with such data, as they are designed to operate on flat, unstructured data. 
On the other hand, GDL methods can directly exploit the data structure, leading to better outcomes \cite{b19}. 

Since its inception, GDL can be applied to many forms of data represented as geometric priors \cite{bronstein2021}. 
Geometric priors encode information about the geometry of the data, such as smoothness, sparsity, or the relationship between the data and other variables, allowing us to work with data of higher dimensionality.
Symmetry is essential in GDL and is often described as invariance and equivariance. 
Invariance is a property of particular mathematical objects that remain unchanged under certain transformations. At the same time, equivariance is a property of certain relations whereby they remain in the same relative position to one another under specific changes \cite{cohen2016, cohen2018}.
For example, consider the graph isomorphism property from \sref{graphtheory}.
Another advantage of GDL is its flexibility toward a broader range of data types and problems. 
This flexibility makes it a powerful tool for solving real-world problems that Euclidean methods cannot.

%% file: input/background/gnn.tex
Similar to traditional deep learning models that use convolutional neural networks (CNNs), GDL utilizes GNNs, which are used to learn representations of graph-structured data, such as social networks, molecules, and computer programs. 
GNNs are very similar to CNNs in that they can extract localized features and compose them to construct representations \cite{lecum1998}. 
As previously mentioned, the critical difference between the two is that GNNs can learn on non-euclidean data and handle data that is not evenly structured, like images or text. 

Training a GNN for graph classification (i.e., determining whether a graph has a characteristic) is relatively simple. There is typically a three-step process that needs to occur: 1) embed each node by performing multiple rounds of message passing, 2) aggregate node embeddings into a unified graph embedding, and then 3) train a final classifier on the graph embedding.
The specific layers selected for a GNN are a choice to be made, and the next section will discuss this.

%% file: input/methodology/intro.tex
The methodology for applying GDL to graph-based design problems will have a multi-label graph classification approach.
This section will discuss this concept, the selected layers, hyperparameters, and metrics used to evaluate the model's performance.

\subsection{Graph Classification}
\input{input/methodology/graph_class}

\subsection{Datasets}
\label{datasets}
\input{input/methodology/datasets}

\subsection{Feature Engineering Through Principal Component Analysis}
\label{feateng}
\input{input/methodology/feature_engineering}

\subsection{The Classification Model Architecture} 
\label{model}
\input{input/methodology/model}

\subsection{Hyperparameters and Options}
\input{input/methodology/hyperparameters}

\subsection{Model Performance Metrics}
\label{metrics}
\input{input/methodology/metrics}

\subsection{Iterative Classification for Graph Down-selection}
\label{sec:iterative-class}

Originally defined in Ref.~\cite{sirico2023}, we will also utilize an iterative classification approach for graph down-selection.
In particular, when we seek binary classification for best graph performance from Sec.~\ref{sec:binary}, classifying based on median performance (once) only reduces the set of $\mathcal{G}_{unknown}$ by approximately half.
This outcome can still result in too many graphs to evaluate, so we might consider iteratively building more focused GDL models based on better threshold values for the ``good'' performance classification.

This approach is outlined in \alref{algorithm}.
With each iteration, the size of $\mathcal{G}^{k}_{known}$ is reduced significantly, progressively lowering the median performance of this set. 
Consequently, the good/bad classification threshold for the GDL model $m^k$ adjusts accordingly.
Then, this new model is used to predict on $\mathcal{G}^{k}_{unknown}$ with those labeled ``Predicted 1'' retained and ``Predicted 0'' discarded as not worth further consideration.

The initial step and key variable to this approach is determining the initial value for $\mathcal{G}^k_{\text{known}}$.
There is not a predefined approach to determining this value, but there are factors that go into it, such as the size of $\mathcal{G}_{all}$, the time it takes to evaluate $\mathcal{G}^k_{\text{known}}$, and the variation in the graph topologies and features. 
If the graph features (graph-level, node, or edge) and graph topology vary widely, the model may need to be trained on a larger number of graphs.

The next important value is the minimum value for the size of $\mathcal{G}_{known}^{n-1}$.
As iterations continue, and we narrow down to the top graphs in the set, $\mathcal{G}_{known}$ is going to continue to shrink as we remove graphs that were predicted as ``bad''.
But, there is a minimum number of graphs required for the model to train and be effective.
Therefore, if we set a minimum value for this set, $|\mathcal{G}_{known}|_{min}$, we can supplement it with predicted 1s from the previously completed iteration.
An initial value for $|\mathcal{G}_{known}|_{min}$ of 100 graphs was chosen, but due to the small dataset size of 2,098 graphs, we found that the model narrowed down fairly quickly to the point where by the fourth iteration, the model was becoming ineffective, and the model only found one graph out of the top 10.
After more tests were conducted, we found that approximately 10\% of the size of $\mathcal{G}_{all}$ works very well, and was used for the iteration experiment in \sref{sec:iterclass}.

\begin{algorithm}[t]
    \caption{Iterative Classification for Graph-Set Based Design Algorithm}
    \begin{algorithmic}[1]        
        \State \textbf{Input:} $|\mathcal{G}_{known}|_{min}$, $n$, $\mathcal{G}_{all}$
        
        \State $k \gets 1$ \Comment{Initialize iteration counter}
        \State Determine $N^k_{known}$
        \State Randomly select $N_{known}$ graphs from $\mathcal{G}_{all}$

        \State Evaluate $\mathcal{G}_{known}$
        
        \State Classify $\mathcal{G}^k_{\text{known}}$ into ``Known 1'' and ``Known 0'' based on $\bar{J}(\mathcal{G}^{k}_{known})$
        
        \While{$k \leq n$} \Comment{Iterate until specified limit}
            
            \State Train GDL model $m^k$ using $\mathcal{G}^k_{known}$ \Comment{Train the model}        
            
            \State $m^k(\mathcal{G}^k_{unknown})$ \Comment{Model predicts on $\mathcal{G}_{unknown}$}
            
            \State Form sets ``Predicted 1'' and ``Predicted 0'' from these predictions

            \State Discard ``Predicted 0'' and ``Known 0'' Classes

            \State Set $\mathcal{G}^{k+1}_{known} \gets$ ``Known 1''
            
            \If{$N^{k+1}_{known} < |\mathcal{G}_{known}|_{min}$} \Comment{Check if known dataset is below the minimum size}
            
                \State Supplement $\mathcal{G}^{k+1}_{known}$ with graphs from ``Predicted 1'' until $N^{k+1}_{known} = |\mathcal{G}_{known}|_{min}$

                \State Divide $\mathcal{G}^{k+1}_{known}$ into ``Known 1'' and ``Known 0'' based on $\bar{J}(\mathcal{G}^k_{known})$

                \State Set $\mathcal{G}^{k+1}_{unknown} \gets$ ``Predicted 1''

            \Else

                \State Divide $\mathcal{G}^{k+1}_{known}$ into ``Known 1'' and ``Known 0'' based on $\bar{J}(\mathcal{G}^{k+1}_{known})$ 

                \State Set $\mathcal{G}^{k+1}_{unknown} \gets$ ``Predicted 1''
                
            \EndIf        
            
            \State $k \gets k + 1$ \Comment{Increment iteration counter}
        \EndWhile
    \end{algorithmic}
    \label{algorithm}
\end{algorithm}

\subsection{Geometric Deep Learning's Roll in Graph Set-Based Design}
\input{input/methodology/sbd2}

%% file: input/methodology/graph_class.tex
This study categorizes entire graphs by assigning them to specific groups based on their characteristics, known as graph classification. 
Specifically, we are working with supervised graph classification. 
Generally, this technique means we have a collection of graphs, denoted as \( \mathcal{G} \), and each graph in this collection, \(G_i \), is tagged with a label, \(L(G_i) \) (or even multiple labels).
The realm of graph classification encompasses a variety of techniques, including Deep Graph Convolutional Neural Networks (DGCNN) \cite{zhang2018}, methods involving hidden layer representations that encapsulate both graph structure and node attributes \cite{b8}, EigenPooling \cite{b12}, and differentiable pooling \cite{ying2019hierarchical}. 
These methods have found application in diverse domains, including areas such as learning molecular fingerprints \cite{duvenaud2015}, text categorization \cite{rousseau2015}, encrypted traffic analysis \cite{shen2021}, and cancer research \cite{hashemi2013}.

Two different scenarios that use graph classification will now be described.

\subsubsection{Multi-Label Classification for Model Compilation and Simulatability}
\label{sec:multi-label}
When we have multiple labels, we can represent them for graph $G_i$ as:
\begin{align}
    \mathbf{L}(G_i) = [l_1, l_2, \ldots, l_{c-1}, l_c]
\end{align}

\noindent where in this label array, each \(l_c\) represents a different aspect or \textit{sub-label}. 
The rationale for adopting a multi-label strategy is to be able to answer several different questions for each graph using a single model instead of requiring multiple models for other questions. 

For the area of interest in this paper, we are interested in answering two particular questions about each graph (see \sref{enum} and \sref{case} for specifics):
\begin{enumerate}
\item Can the graph be compiled (i.e., can a physics-based model be automatically constructed from a given graph)?
\item  Is the graph capable of being simulated (i.e., given a complied model, are there valid simulation outcomes)?   
\end{enumerate}

\noindent Based on the answers to these questions, we derive four possible categories for the graphs: graphs that will not compile, graphs that will compile, graphs that will not simulate, and graphs that will simulate.
Now, each graph will have the label array \(\mathbf{L}(G_i) = [l_1, l_2, l_3, l_4]\). 
The subset \([l_1, l_2]\) corresponds to the graph's ability to compile (or not), and \([l_3, l_4]\) relates to its ability to simulate (or not).
The three classification options are then:
\begin{align}
\mathbf{L}(G_i) = 
\begin{cases}
[0, 1, 0, 1] & \text{if $G_i$ will compile and simulate} \\
[0, 1, 1, 0] & \text{if $G_i$ will compile but not simulate} \\
[1, 0, 1, 0] & \text{if $G_i$ will not compile}
\end{cases}
\end{align}

\subsubsection{Binary Classification for Best Graph Performance}
\label{sec:binary}

In this scenario, some median performance value $\bar{J}$ is a benchmark for categorizing our graphs into two binary classes, where a smaller value $J(G_i)$ is considered better or more desirable. 
Graphs falling in the lower half of the observed performance values (i.e., $J(G_i) \leq \bar{J}$) are, therefore, labeled as ``good'' and assigned the label 1, while those in the upper half (i.e., $J(G_i) > \bar{J}$) are considered ``bad'' and assigned the label 0.
This is discussed further in \sref{datasets}.

Here, the chosen approach, favoring classification over regression, aligns more closely with the designer's objectives in the early stages of conceptual design, notably in facilitating flexible down-selection. 
The primary aim here is not to pinpoint a single optimal graph but to identify a set of ``good'' or promising graphs for further detailed analysis and optimization.
This need arises because, at this stage, there is often a reliance on assumptions in modeling and other aspects of graph representation and design, which necessitates a higher-fidelity examination of the shortlisted graphs.
Moreover, the performance values \( J(G_i) \) in the case study exhibit significant variations, and some values are not well-defined due to the graph not simulating (or compiling).
Consequently, the focus is on categorizing graphs into particular groups rather than assigning precise values to each. 

%% file: input/methodology/datasets.tex
For both scenarios described above, we will investigate when only a subset of outcomes (e.g., labels $\mathbf{L}(G_i)$ or performance value $J(G_i)$) are known, dividing the graphs into two distinct sets:
\begin{align}
\mathcal{G} \equiv \mathcal{G}_{all} = \mathcal{G}_{known} \cup \mathcal{G}_{unknown}
\end{align}

Here, $\mathcal{G}_{known}$ represents the set of graphs with known outcomes, while $\mathcal{G}_{unknown}$ includes graphs with unknown outcomes.
We'll denote the sizes of these sets as $|\mathcal{G}_{all}| = N_{all}$, $|\mathcal{G}_{known}| = N_{known}$, and $|\mathcal{G}_{unknown}| = N_{unknown}$. It's also essential to emphasize that this dataset structure differs from other machine learning datasets, as there is a finite and known set of potential inputs (i.e., the list of graphs). The primary objective is to develop accurate models when $N_{known} \ll N_{unknown}$.

Following the typical learning approach, we will further split $\mathcal{G}_{known}$ into two subsets:
\begin{align}
\mathcal{G}_{known} = \mathcal{G}_{training} \cup \mathcal{G}_{validation}
\end{align}

\noindent
Here, $\mathcal{G}_{training}$ serves as the training dataset, while $\mathcal{G}_{validation}$ is designated as the validation dataset. The model is trained using $\mathcal{G}_{training}$, and the fitted model subsequently predicts responses for the observations in $\mathcal{G}_{validation}$ \cite{b46}.
For instance, following each iteration, the model updates its weights and evaluates them using the validation set. 
This process aids in assessing the model's performance and facilitates necessary adjustments. 
Once the model is fully developed, it is tested on the unknown set (as we do have their outcomes in this study) to gauge its effectiveness on previously unseen data.

In the first scenario in \sref{sec:multi-label}, the multi-label classification has been determined for the graphs within the set \( \mathcal{G}_{known} \). 
Conversely, in the second scenario in \sref{sec:binary}, the binary classification of graphs in \( \mathcal{G}_{known} \) hinges on the median performance value or possibly another criterion derived from the known data. Given that performance values \( J(G_i) \) are available only for \( \mathcal{G}_{known} \) and not for the entire set \( \mathcal{G}_{all} \), the median value used for initial labeling might deviate from what would be the case if all performance values in \( \mathcal{G}_{all} \) were known. Nonetheless, as the relative size of \( \mathcal{G}_{known} \) grows, this discrepancy is expected to diminish.

%% file: input/methodology/feature_engineering.tex
In this context, we explore augmenting the feature set \(\mathbf{X}\) with additional graph-based attributes (extending beyond the existing vertex labels such as HEX, BSO, and BSI). 
Incorporating these supplementary features can enhance the model's capacity to account for the variance within the training data, a practice commonly referred to as feature engineering and feature selection.
The underlying hypothesis is that the added attribute will potentially enhance the model's performance while requiring similar training epochs.
The potential issue is that several graph candidates' underlying structures may be very similar, resulting in identical graph-based features.
Therefore, a technique known as \textit{Principle Component Analysis (PCA)} is used.

\subsubsection{Principle Component Analysis (PCA)}
Principal Component Analysis, commonly known as PCA, is a technique used to reduce complex data sets' dimensions. 
PCA projects the data's most critical features onto a smaller set of dimensions known as principal components. 
This makes analyzing and interpreting the data easier and is therefore valuable across various fields. 
Graph-based data sets can also offer an advantage in extracting features directly from the graphs themselves, especially when dealing with a dataset that does not have a lot of base features\cite{scikit-learn}.
The following features were extracted from the TMS dataset and applied to each graph.

\begin{enumerate}
    \item Harmonic Centrality
    calculated as the sum of the reciprocals of the shortest path distances \(d\) from every other node to a specific node \(u\)\cite{boldi2013axioms}, mathematically represented as:
    \[
        C(u) = \sum_{v \neq u} \frac{1}{d(v,u)}
    \]
    
    \item Betweenness Centrality
    quantifies the extent to which a node lies on the shortest paths between other node pairs. It is defined as:
    \[
    c_B(v) = \sum_{s,t \in V} \frac{\sigma(s,t | v)}{\sigma(s,t)}
    \]
    where \(V\) denotes the set of nodes, \(\sigma(s,t)\) is the total number of shortest paths from node \(s\) to node \(t\), and \(\sigma(s,t | v)\) counts those paths passing through node \(v\), excluding direct paths between \(s\) and \(t\).\cite{b61}
    
    \item Eigenvector Centrality assesses a node's influence based on the centrality of its neighbors, implying that a node is considered vital if it is connected to other significant nodes. 
    The centrality for node \(v\) is the \(i\)-th normalized element of vector \(\mathbf{v}\), determined by:
    \[
    \mathbf{A}\mathbf{v} = \lambda \mathbf{v}
    \]
    Here, \(\mathbf{A}\) is the adjacency matrix, and \(\lambda\) represents the largest eigenvalue.\cite{b60}
    
    \item Spectral Radius
    is the maximum absolute value among the eigenvalues (\(|\lambda_1|, \dots, |\lambda_n|\)) of a graph's adjacency matrix:
    \[
        \rho(\mathbf{A}) = \max\{|\lambda_1|, \dots, |\lambda_n|\}
    \]
    The spectral radius provides insights into the graph's connectivity and robustness, indicating that a high value suggests a well-connected structure.\cite{Stevanović2018}
\end{enumerate}

\begin{figure}[t]
    \centering
    \includegraphics[scale=0.6]{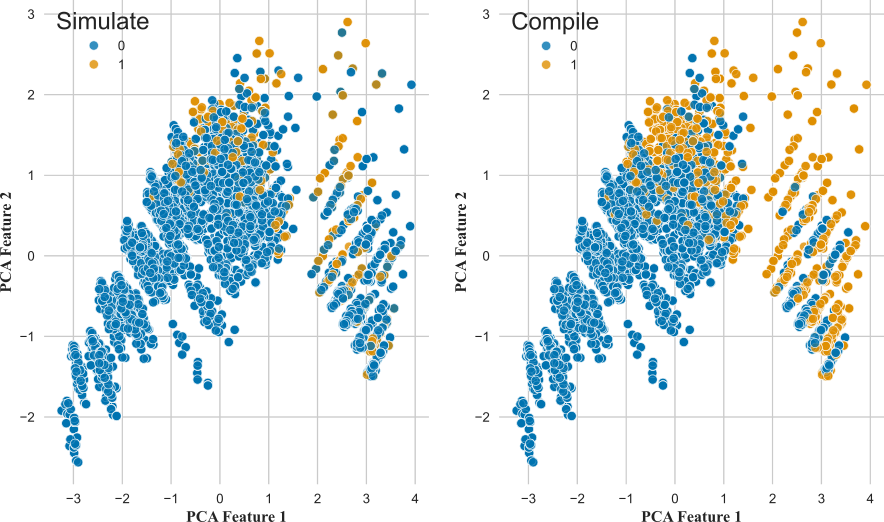}
    \caption{The Principle Component Analysis features displayed for both the simulate and compile classifications.}
    \label{fig:pca}
\end{figure}

\noindent These four features were passed through a PCA model, and the results are shown in \fref{fig:pca}.
You can see the distinction between the two classification sets.
In \sref{sec:featureeng}, we apply these features to the dataset and compare it to a model that did not use any additional features to demonstrate the benefits of adding new features to the dataset.

%% file: input/methodology/model.tex
In this study, the model comprises five layers: namely, three graph convolutional layers \cite{morris2018}, a mean pooling layer, and a linear layer for the ultimate output. To achieve the objective of graph classification, the model follows a three-step process: 1) employs the convolutional layers to encode individual nodes by exchanging messages, 2) aggregates the node embeddings to form a graph representation, and 3) utilizes the graph representation to train the classification layer.

\subsubsection{Graph Convolutional Layer} The Graph Convolutional Layer (GCN) operates to derive the output features $\textbf{X}'$ following this formula:
\begin{align}
\textbf{X}'_i = \textbf{W}_1\textbf{X}_i + \textbf{W}_2 \sum_{j\in \mathrm{N}(i)}e_{j,i}\cdot \textbf{X}_j
\label{gcn}
\end{align}

\noindent where $\textbf{X}$ represents the input feature for each node and $e_{j,i}$ denotes the edge weight from source node $j$ to target node $i$ \cite{morris2018}.
Looking at \eref{digrapheq}, the last right column represents the features matrix for that particular graph.
The weights $(\textbf{W}_1, \textbf{W}_2)$ adapt during training iterations.

Each GCN layer also employs the Rectified Linear Unit (ReLU) activation function $f(x) = \max(0, x)$,
which returns 0 for any negative input and returns the input value for positive inputs. Finally, the result from \eref{gcn} is passed through the activation function:
\begin{align}
\textbf{X}_{i+1} = f(\textbf{X}_i) 
\label{gcn2}
\end{align}

\begin{figure}[t]
\centering
\includegraphics[scale=0.7]{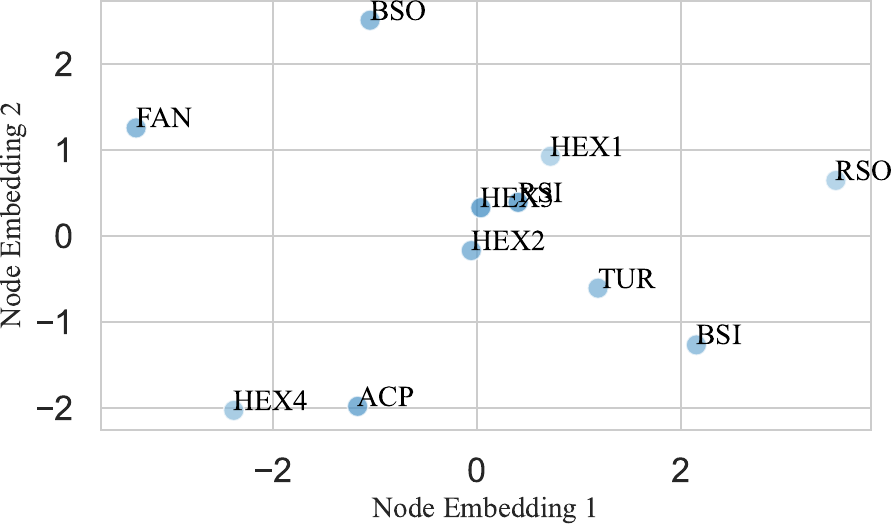}
\caption{Computed node embeddings for the ACM graph in \fref{acmgraph}.}
\label{node_embed}
\end{figure}

\noindent This step assists in obtaining localized node embeddings.
Node embeddings involve translating the nodes or vertices of a graph into an $N$-dimensional numeric space, resulting in numerical vectors that encapsulate the graph's structure. 
This is exemplified through the node embeddings obtained for the ACM graph shown in \fref{acmgraph} and depicted in \fref{node_embed}.
In this illustration, it is evident that nodes HEX1, HEX2, and HEX3 (which are the top three HEX nodes in \fref{acmgraph}, respectively) along with RSI, have been identified as similar and, consequently, are clustered together.

\subsubsection{Global Mean Pooling Layer} The subsequent layer is the Global Mean Pooling layer. This layer processes the final output from the GCN layers. 
It produces an output vector $\textbf{r}$ by computing the mean of the node features $\textbf{X}$, encompassing node embeddings for the entire graph, spanning all nodes within the graph $N$. 
This procedure generates a comprehensive graph embedding:
\begin{align}
\textbf{r}_i = \frac{1}{N_i}\sum^{N_i}_{n=1} \textbf{X}_n
\label{gmp}
\end{align}

\subsubsection{Linear Layer} The last layer in the model is the classification layer. 
It receives the output from the mean pooling layer as its input and carries out a linear transformation.
Nevertheless, just before this layer, a ``dropout'' operation is employed, which stochastically sets certain elements of the input tensor to zero with a specified probability, utilizing samples from a Bernoulli distribution, as introduced in \rref{hinton2012}.

%% file: input/methodology/hyperparameters.tex
Several adjustable factors can be fine-tuned within the chosen model architecture to ensure adequate training on the provided data. For a comprehensive understanding of hyperparameters and their optimization, please consult the following references \cite{b44, b45, b46}.

\begin{table}[h!]
\centering
\caption{Hyperparameters and their descriptions.}
\begin{tabularx}{400pt}{rX}
\hline \hline
\textbf{Parameter} & \textbf{Description} \\
\hline
Learning Rate (LR) & The model’s learning rate is a crucial hyperparameter represented by a positive scalar value, typically between 0.0 and 1.0. It determines the step size at each epoch during weight adjustment. A learning rate of 0.001 is used in these case studies. \\
\hline
Number of Epochs & An epoch is a full pass through the entire training dataset. The number of epochs is variable and can be hundreds to thousands, ensuring the model trains until error minimization. \\
\hline
Optimization Algorithm & The Adaptive Movement Estimation (ADAM) optimization algorithm is used. It is a stochastic technique that calculates individual learning rates for different parameters. \\
\hline
Loss Function & The loss function employed is the cross-entropy loss, which is suitable for classification problems and measures the dissimilarity between two probability distributions. \\
\hline
Batch Size & Batch Size refers to the number of data points processed before the model’s weights are updated. Mini-batching allows for scalability and involves processing an intermediate-sized group of data points. \\
\hline \hline
\end{tabularx}

\label{table:1}
\end{table}

%% file: input/methodology/metrics.tex
\subsubsection{Confusion Matrix}
\begin{figure}[ht]
    \centering
    \includegraphics[width=0.3\textwidth]{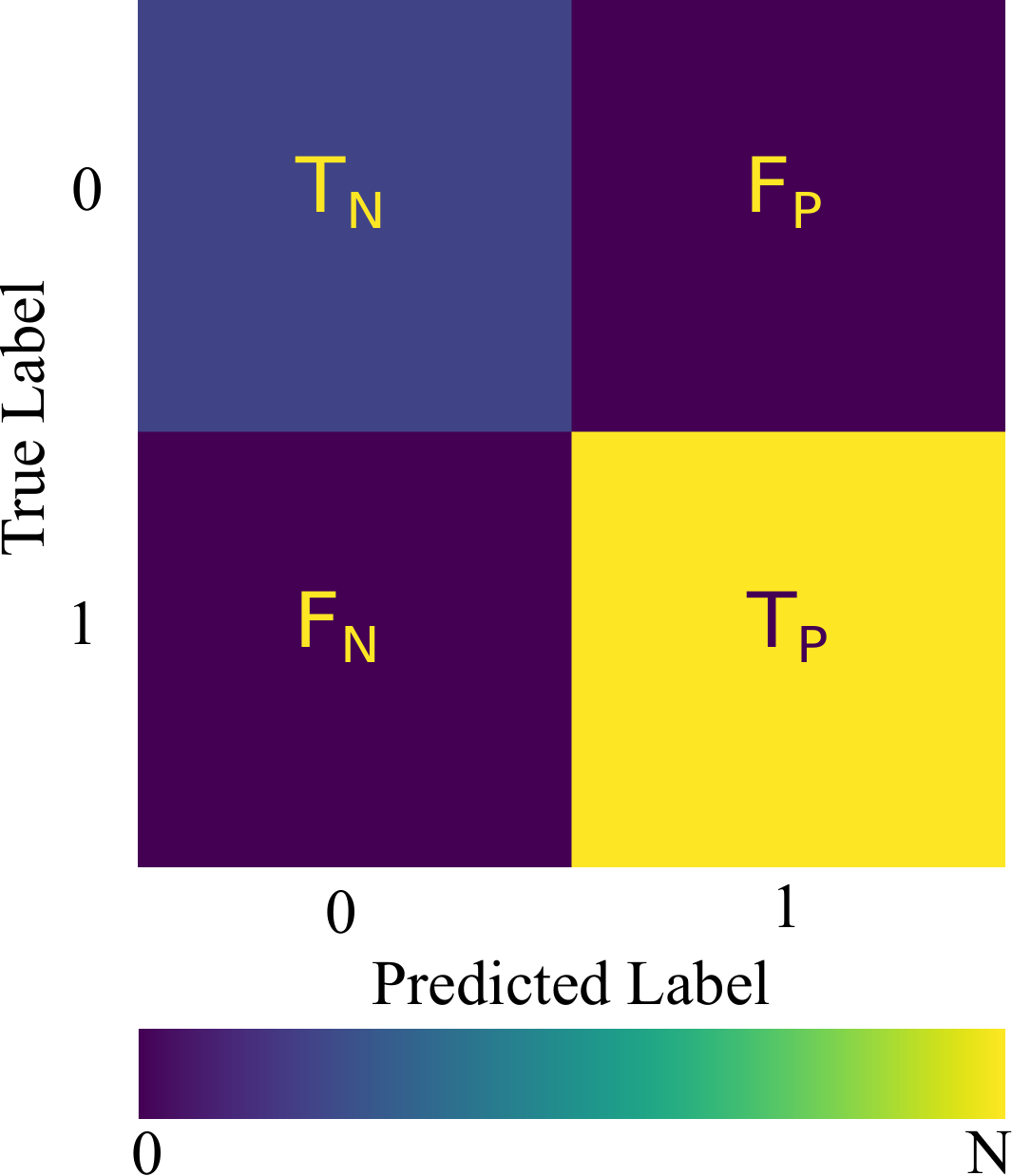}
    \caption{An example confusion matrix.}
    \label{fig:cm}
\end{figure}

~A confusion matrix, illustrated in \fref{fig:cm}, is a tool used in classification to visualize the performance of a model. 
It is a two-dimensional matrix where the columns represent the instances as classified by the model, and the rows represent the instances as they are in the true dataset, as described in \cite{simske2013}. 
In the context of a binary classifier, the matrix is divided into four parts: the top left quadrant shows the True Positives (\(T_P\)), which are the data points correctly identified as ones. 
The top right quadrant represents the False Positives (\(F_P\)), where the data points are incorrectly labeled as ones when they are zeros. 
Conversely, the bottom left quadrant contains the False Negatives (\(F_N\)), indicating the points mistakenly classified as zeros but true ones. 
Finally, the bottom right quadrant is for the True Negatives (\(T_N\)), which correctly identifies the data points as zeros. 
The counts in each section of this matrix form the basis for many subsequent performance metrics and are shown in \tref{tab:metrics}.

\begin{table}[t]
\centering
\caption{The metrics used from the extrapolated information from the confusion matrix.}
\begin{tabularx}{\textwidth}{>{\raggedleft}p{0.19\textwidth}cX}
\hline\hline
\textbf{Metric} & \textbf{Formula} & \textbf{Definition}\\
\hline
Accuracy (ACC) & $\displaystyle\frac{T_P + T_N}{N}$ & Accuracy is the number of correct predictions divided by the total number of predictions ($N$)\\ 
$Precision$ & $\displaystyle\frac{T_P}{T_P+F_P}$ & Measures the accuracy of positive predictions\\
$Recall$ & $\displaystyle\frac{T_P}{T_P+F_N}$ & Indicates the proportion of actual positive classifications were identified correctly\\
F1 & $\displaystyle2 \cdot \frac{\textit{Precision} \cdot \textit{Recall}}{\textit{Precision} + \textit{Recall}}$ & The harmonic mean between precision and recall\\
Matthews Correlation Coefficient (MCC) & $\displaystyle\frac{T_P \cdot T_N - F_P \cdot F_N}{\sqrt{(T_P + F_P)(T_P+F_N)(T_N+F_P)(T_N+F_N)}}$ & Ranges $[-1,1]$  and produces a high score if the model obtained good results in \textit{all} boxes of the confusion matrix \cite{jurman2012, chicco2017} \\
\hline\hline
\end{tabularx}
\label{tab:metrics}
\end{table}

\subsubsection{Receiver Operating Characteristic (ROC) Curve}

Another metric extremely useful in classification models is the Area Under the Receiver Operating Characteristic (ROC) curve.
The ROC curve represents a graphical depiction of a classification model's performance using $Recall$ vs. False Positive Rate $F_P/(F_P + T_N)$ across various threshold levels.
The total area under the ROC curve (AUC) ranges from point $(0,0)$ to $(1,1)$.
Ranging from 0 to 1, a model with an AUC of 1.0 predicted everything 100\% correctly, while a model with an AUC of 0 was completely wrong in its predictions.

%% file: input/methodology/sbd2.tex
IC-GSBD exemplifies the innovative application of SBD principles in conjunction with graph-based design and GDL. 
This integration leverages the SBD methodology to enhance the design process significantly by \textbf{broadening the exploration of design alternatives} utilizing graph-based models to represent various configurations, embracing the SBD principle of extensive exploration.
Generating a \textbf{more informed narrowing and selection process} by harnessing the capabilities of GDL, IC-GSBD enriches the SBD framework, enabling an efficient evaluation and down-selection of design alternatives. 
The proficiency of GDL in processing graph representations expedites the assessment of each design's feasibility and performance, streamlining the refinement process \cite{bronstein2021}.
IC-GSBD also \textbf{enhances the decision-making process} with Computational Efficiency by embodying the systematic and iterative exploration central to SBD, augmented by the computational efficiency and analytical depth of GDL. 
This combination facilitates informed decision-making and may reduce the number of design iterations required \cite{sirico2023}.

\begin{figure}[t]
    \centering
    \begin{subfigure}[c]{0.48\textwidth}
        \begin{tikzpicture}[node distance=1.8cm, scale=0.65, every node/.style={transform shape}]
            \node (start) [process] {Define Requirements and Constraints};
            \node (generate) [dygide, below of=start, yshift=-0.3cm] {Generate Set of Graphs};
            \node (evaluate) [dygide, below of=generate] {Evaluate a Fraction of the Dataset};  
            \node (machine) [dygide, below of=evaluate, yshift=-0.2cm] {Machine Learning Process};
            \node (decision) [decision, below of=machine, yshift=-1cm] {Does the design meet stakeholder needs?};
            \node (refine) [process, left of=decision, xshift=-3.5cm] {Iteratively Refine Alternatives};
            \node (final) [process, below of=decision, yshift=-1.5cm] {Finalize Design and Develop Detailed Specifications};

            \node[right] at (start.north east) {1};
            \node[right] at (generate.north east) {2};
            \node[right] at (evaluate.north east) {3};
            \node[right] at (machine.north east) {4};
            \node[above right] at (decision.north east) {5};
            \node[left] at (refine.north west) {5a};
            \node[right] at (final.north east) {5b};
        
            \draw [arrow] (start) -- (generate);
            \draw [arrow] (generate) -- (evaluate);
            \draw [arrow] (evaluate) -- (machine);
            \draw [arrow] (machine) -- (decision);
            \draw [arrow] (decision) -- node[anchor=east] {yes} (final);
            \draw [arrow] (decision) -- node[anchor=south] {no} (refine);
            \draw [arrow] (refine) |- (evaluate);
        \end{tikzpicture}
        \caption{The basic workflow for Set-Based Design.}
        \label{fig:gsbdwfa}
    \end{subfigure}    
    \hfill
    \begin{subfigure}[c]{0.48\textwidth}
        \centering
        \vfill
        \begin{tikzpicture}[node distance=1.8cm, scale=0.65, every node/.style={transform shape}]
            \node (train) [dygide] {Train GDL Model};
            \node (predict) [automate, below of=train] {Predict the \textit{unknown} Set};
            \node (remove) [automate, below of=predict] {Remove ``Bad'' Predictions};
            \node (reclass) [automate, below of=remove] {Reclassify New ``Known'' Set};

            \node[right] at (train.north east) {4.1};
            \node[right] at (predict.north east) {4.2};
            \node[right] at (remove.north east) {4.3};
            \node[right] at (reclass.north east) {4.4};

            \draw [arrow] (train) -- (predict);
            \draw [arrow] (predict) -- (remove);
            \draw [arrow] (remove) -- (reclass);
            \draw [arrow] (reclass) -- ([xshift=1cm]reclass.east) |- ([xshift=1cm]train.east) -- (train);
        \end{tikzpicture}
        \vfill
        \caption{An exploded view of the machine learning process.}
        \label{fig:gsbdml}
    \end{subfigure}
    \caption{The IC-GSDB workflow where purple indicates the changes to the former SBD process and red indicates automated processes.}
    \label{fig:gsbdwf}
\end{figure}
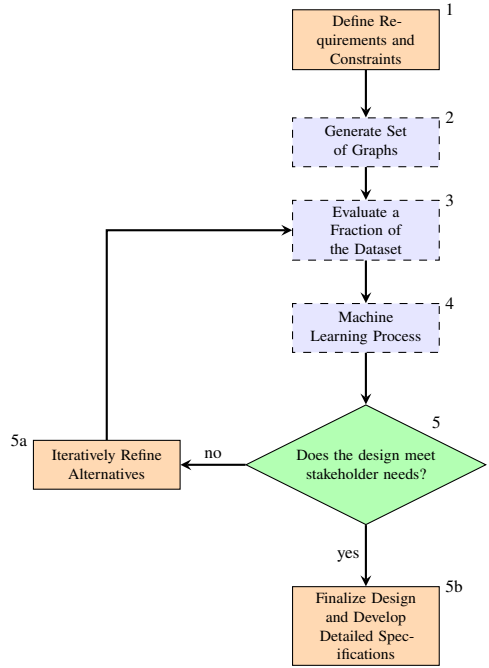
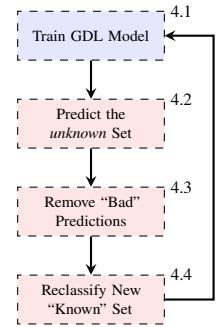

The IC-GSBD framework utilizes an SBD approach to categorize and refine graph-based TMS designs. 
As illustrated in \fref{fig:sbdflow}, the process begins methodically with defining requirements and generating design alternatives. 
These alternatives are manually evaluated and narrowed down to determine the optimal design solution. 
In \fref{fig:gsbdwf}, we integrate GDL and iterative classification into the foundational SBD workflow.

One significant advantage of this revised workflow, as depicted in Figure \ref{fig:gsbdwfa}, is that it does not introduce additional steps to the existing process. 
Furthermore, \fref{fig:gsbdml} presents an exploded view of the machine learning process box, highlighting the incorporation of automated processes that refine the alternatives efficiently.

This enhanced methodology fits seamlessly within the traditional SBD framework and enhances it by streamlining the management of computational complexity associated with expansive design explorations. 
It provides a robust mechanism for evaluating and predicting the performance and feasibility of a broad range of design possibilities, represented as graphs, with remarkable efficiency and effectiveness.

%% file: input/results/intro.tex

\subsection{Dataset Overview}
\label{case}

The dataset used in this study is from \rref{herber2020}, where an aircraft TMS is tasked with controlling the temperatures of two hypothetical loads: a flight control heat load and a radar heat load. 
The designated target temperatures are 343K for the flight control and 300K for the radar. 
A multi-objective utility function is employed to rank and select the most suitable architectural designs. 
This function considers various goals, including keeping thermal loads within specified temperature ranges. The utility function is defined as follows:
\begin{eqnarray} \label{eq:J}
    \text{minimize: } J &=& w_1 J_1 + w_2 J_2 + w_3 J_3 + w_4 J_4 + w_5 J_5\\
     &=& w_1 (\bar{T}_{HLF} - 343K) + w_2(\bar{T}_{HLR} - 300K) + w_3 C_{total} + w_4\dot{m}_{BSI} + w_5\dot{m}_{RSI}
\end{eqnarray}

\noindent where \(\bar{T}_{HLF}\) represents the average temperature of the flight control heat load, \(\bar{T}_{HLR}\) is for the radar heat load, \(C_{total}\) denotes the total heat capacity of the heat exchangers, and \(\dot{m}_{BSI}\) and \(\dot{m}_{RSI}\) are the mass flow rates of the bleed air and ram air, respectively. Nominal weights have been established to provide a singular value for ranking different architectural solutions.

Employing the techniques outlined in \sref{enum}, a total of 32,612 potential architectures, represented as graphs, were evaluated.
Of the considered graphs, 5,585 were successfully compiled, and among these, 2,098 could be simulated.
Simulatability and best performance $J(G_i)$ were determined by trying to simulate compiled models using 200 parameter sets.
If at least one parameter set produced a valid result, then it is considered simulatable, and the lowest $J$ value from the valid results as assigned to $J(G_i)$.

\input{input/results/examples}

\subsection{A Model for Multi-Label Classification for Model Compilation and Simulatability}


We first focus on developing a GDL model for the first scenario from \sref{sec:multi-label} as understanding if a graph will compile and simulate is the first critical step in their evaluation.

\input{input/results/exp1}
\label{exp1}

\input{input/results/exp2}

\label{exp2}

%% file: input/results/examples.tex
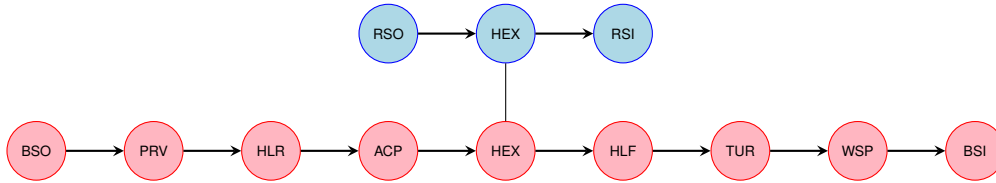
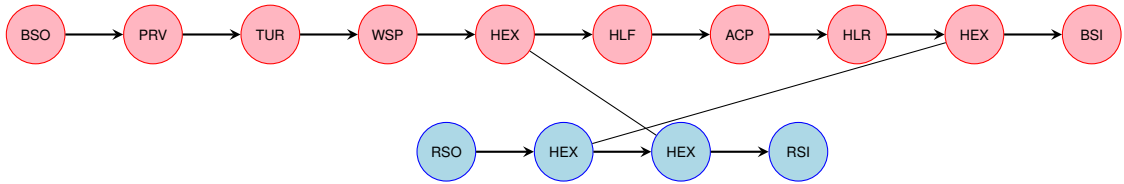
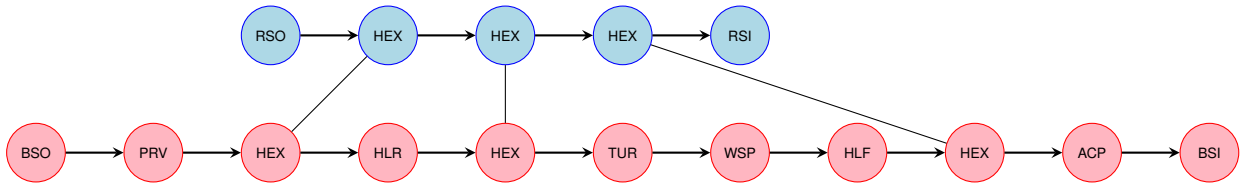
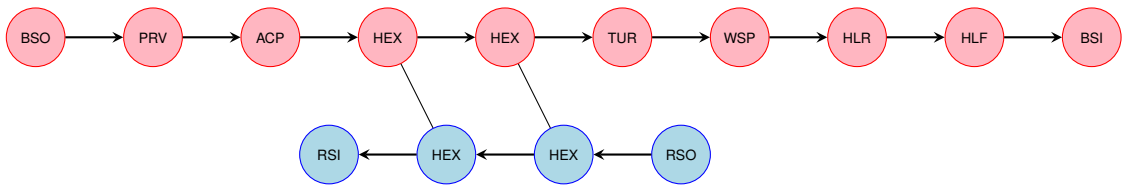
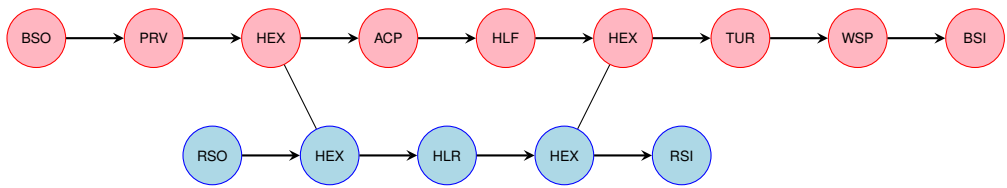
\begin{figure}[p]
\newcommand{\tmsscale}{0.78}
    \centering
    \begin{subfigure}[c]{\textwidth}
        \centering
        \input{input/results/acm_plots}
    \caption{A \textit{simulatable} 17-node directed graph representing a TMS with three heat exchangers and $J=469.9K$}
    \label{fig:tmsgraph2}
    \end{subfigure}
    \vspace{1em}
    \begin{subfigure}[c]{\textwidth}
        \centering
        \input{input/results/acm2}
        \caption{A \textit{simulatable} 12-node directed graph representing a TMS with a single heat exchanger and $J=719.04K$}
        \label{fig:tmsgraph3}
    \end{subfigure}
    \vspace{1em}
    \begin{subfigure}[c]{\textwidth}
        \centering
        \input{input/results/tms3}
        \caption{A 14-node directed graph representing a TMS that \textit{did not compile}}
        \label{fig:tmsgraph4}
    \end{subfigure}
    \vspace{1em}
    \begin{subfigure}[c]{\textwidth}
        \centering
        \input{input/results/tms4}
        \caption{A 16-node graph representing a TMS that \textit{compiled}, but did not \textit{simulate}}
        \label{fig:tmsgraph5}
    \end{subfigure}
    \vspace{1em}
    \begin{subfigure}[c]{\textwidth}
        \centering
        \input{input/results/tms5}
        \caption{A 14-node directed graph representing a TMS that had the lowest value of $J=-167.82 K$}    
        \label{fig:tmsgraph6}
    \end{subfigure}
    \vspace{1em}
    \begin{subfigure}[c]{\textwidth}
        \centering
        \input{input/results/tms6}
        \caption{A 14-node directed graph representing a TMS that had the highest value of $J=5.33 \times 10^3 K$}
        \label{fig:tmsgraph7}
    \end{subfigure}
    \caption{Two more examples of TMSs  with varying numbers of heat exchangers but also exhibiting other TMS components}
    \label{fig:tms_ex}
    
\end{figure}

%% file: input/results/acm_plots.tex
\begin{tikzpicture}[scale=\tmsscale, transform shape, minimum size=1cm]
    \centering
    \node[circle, draw=red, fill=lightred, minimum size=1cm] (bso) at (-10,0) {$\xy{BSO}$};
    \node[circle, draw=red, fill=lightred, minimum size=1cm] (prv) at (-8,0) {$\xy{PRV}$};
    \node[circle, draw=red, fill=lightred, minimum size=1cm] (acp) at (-6,0) {$\xy{ACP}$};
    \node[circle, draw=red, fill=lightred, minimum size=1cm] (hex1) at (-4,0) {$\xy{HEX}$};
    \node[circle, draw=red, fill=lightred, minimum size=1cm] (hex2) at (-2,0) {$\xy{HEX}$};
    \node[circle, draw=red, fill=lightred, minimum size=1cm] (hex3) at (0,0) {$\xy{HEX}$};
    \node[circle, draw=red, fill=lightred, minimum size=1cm] (tur1) at (2,0) {$\xy{TUR}$};
    \node[circle, draw=red, fill=lightred, minimum size=1cm] (hlf) at (4,0) {$\xy{HLF}$};
    \node[circle, draw=red, fill=lightred, minimum size=1cm] (tur2) at (6,0) {$\xy{TUR}$};
    \node[circle, draw=red, fill=lightred, minimum size=1cm] (wsp) at (8,0) {$\xy{WSP}$};
    \node[circle, draw=red, fill=lightred, minimum size=1cm] (bsi) at (10,0) {$\xy{BSI}$};

    \node[circle, draw=blue, fill=lightblue, minimum size=1cm] (rsi) at (-6,-2) {$\xy{RSI}$};
    \node[circle, draw=blue, fill=lightblue, minimum size=1cm] (hex4) at (-4,-2) {$\xy{HEX}$};
    \node[circle, draw=blue, fill=lightblue, minimum size=1cm] (hex5) at (-2,-2) {$\xy{HEX}$};
    \node[circle, draw=blue, fill=lightblue, minimum size=1cm] (hex6) at (0,-2) {$\xy{HEX}$};
    \node[circle, draw=blue, fill=lightblue, minimum size=1cm] (hlr) at (2,-2) {$\xy{HLR}$};
    \node[circle, draw=blue, fill=lightblue, minimum size=1cm] (rso) at (4,-2) {$\xy{RSO}$};

    \draw[arrow] (bso) -- (prv);
    \draw[arrow] (prv) -- (acp);
    \draw[arrow] (acp) -- (hex1);
    \draw[arrow] (hex1) -- (hex2);
    \draw[arrow] (hex2) -- (hex3);
    \draw[arrow] (hex3) -- (tur1);
    \draw[arrow] (tur1) -- (hlf);
    \draw[arrow] (hlf) -- (tur2);
    \draw[arrow] (tur2) -- (wsp);
    \draw[arrow] (wsp) -- (bsi);

    \draw (hex1) -- (hex4);
    \draw (hex2) -- (hex5);
    \draw (hex3) -- (hex6);

    \draw[arrow] (rso) -- (hlr);
    \draw[arrow] (hlr) -- (hex6);
    \draw[arrow] (hex6) -- (hex5);
    \draw[arrow] (hex5) -- (hex4);
    \draw[arrow] (hex4) -- (rsi);
    
\end{tikzpicture}

%% file: input/results/acm2.tex
\begin{tikzpicture}[scale=\tmsscale, transform shape, minimum size=1cm]
    \centering
    \node[circle, draw=red, fill=lightred, minimum size=1cm] (bso) at (-8,0) {$\xy{BSO}$};
    \node[circle, draw=red, fill=lightred, minimum size=1cm] (prv) at (-6,0) {$\xy{PRV}$};
    \node[circle, draw=red, fill=lightred, minimum size=1cm] (hlr) at (-4,0) {$\xy{HLR}$};
    \node[circle, draw=red, fill=lightred, minimum size=1cm] (acp) at (-2,0) {$\xy{ACP}$};
    \node[circle, draw=red, fill=lightred, minimum size=1cm] (hex1) at (0,0) {$\xy{HEX}$};
    \node[circle, draw=red, fill=lightred, minimum size=1cm] (hlf) at (2,0) {$\xy{HLF}$};
    \node[circle, draw=red, fill=lightred, minimum size=1cm] (tur) at (4,0) {$\xy{TUR}$};
    \node[circle, draw=red, fill=lightred, minimum size=1cm] (wsp) at (6,0) {$\xy{WSP}$};
    \node[circle, draw=red, fill=lightred, minimum size=1cm] (bsi) at (8,0) {$\xy{BSI}$};

    \node[circle, draw=blue, fill=lightblue, minimum size=1cm] (rso) at (-2,2) {$\xy{RSO}$};
    \node[circle, draw=blue, fill=lightblue, minimum size=1cm] (hex2) at (0,2) {$\xy{HEX}$};
    \node[circle, draw=blue, fill=lightblue, minimum size=1cm] (rsi) at (2,2) {$\xy{RSI}$};

    \draw[arrow] (bso) -- (prv);
    \draw[arrow] (prv) -- (hlr);
    \draw[arrow] (hlr) -- (acp);
    \draw[arrow] (acp) -- (hex1);
    \draw[arrow] (hex1) -- (hlf);
    \draw[arrow] (hlf) -- (tur);
    \draw[arrow] (tur) -- (wsp);
    \draw[arrow] (wsp) -- (bsi);

    \draw (hex1) -- (hex2);

    \draw[arrow] (rso) -- (hex2);
    \draw[arrow] (hex2) -- (rsi);

\end{tikzpicture}

%% file: input/results/tms3.tex
\begin{tikzpicture}[scale=\tmsscale, transform shape, minimum size=1cm]
    \node[circle, draw=blue, fill=lightblue, minimum size=1cm] (rsi) at (7,-2) {$\xy{RSI}$};
    \node[circle, draw=blue, fill=lightblue, minimum size=1cm] (hex1) at (5,-2) {$\xy{HEX}$};
    \node[circle, draw=red, fill=lightred, minimum size=1cm] (hex2) at (2,0) {$\xy{HEX}$};
    \node[circle, draw=red, fill=lightred, minimum size=1cm] (wsp) at (0,0) {$\xy{WSP}$};
    \node[circle, draw=red, fill=lightred, minimum size=1cm] (tur) at (-2,0) {$\xy{TUR}$};
    \node[circle, draw=red, fill=lightred, minimum size=1cm] (prv) at (-4,0) {$\xy{PRV}$};
    \node[circle, draw=red, fill=lightred, minimum size=1cm] (bso) at (-6,0) {$\xy{BSO}$};
    \node[circle, draw=blue, fill=lightblue, minimum size=1cm] (rso) at (1,-2) {$\xy{RSO}$};
    \node[circle, draw=blue, fill=lightblue, minimum size=1cm] (hex3) at (3,-2) {$\xy{HEX}$};
    \node[circle, draw=red, fill=lightred, minimum size=1cm] (hex4) at (10,0) {$\xy{HEX}$};
    \node[circle, draw=red, fill=lightred, minimum size=1cm] (bsi) at (12,0) {$\xy{BSI}$};
    \node[circle, draw=red, fill=lightred, minimum size=1cm] (hlr) at (8,0) {$\xy{HLR}$};
    \node[circle, draw=red, fill=lightred, minimum size=1cm] (acp) at (6,0) {$\xy{ACP}$};
    \node[circle, draw=red, fill=lightred, minimum size=1cm] (hlf) at (4,0) {$\xy{HLF}$};
    
    \draw[arrow] (hex1) -- (rsi);
    \draw (hex1) -- (hex2);
    \draw[arrow] (wsp) -- (hex2);
    \draw[arrow] (tur) -- (wsp);
    \draw[arrow] (prv) -- (tur);
    \draw[arrow] (bso) -- (prv);
    \draw[arrow] (rso) -- (hex3);
    \draw[arrow] (hex3) -- (hex1);
    \draw (hex3) -- (hex4);
    \draw[arrow] (hex4) -- (bsi);
    \draw[arrow] (hlr) -- (hex4);
    \draw[arrow] (acp) -- (hlr);
    \draw[arrow] (hlf) -- (acp);
    \draw[arrow] (hex2) -- (hlf);
    
\end{tikzpicture}

%% file: input/results/tms4.tex
\begin{tikzpicture}[scale=\tmsscale, transform shape]
    \node[circle, draw=blue, fill=lightblue, minimum size=1cm] (rso) at (0,0) {$\xy{RSO}$};
    \node[circle, draw=blue, fill=lightblue, minimum size=1cm] (hex1) at (2,0) {$\xy{HEX}$};    
    \node[circle, draw=blue, fill=lightblue, minimum size=1cm] (hex2) at (4,0) {$\xy{HEX}$};
    \node[circle, draw=blue, fill=lightblue, minimum size=1cm] (hex3) at (6,0) {$\xy{HEX}$};
    \node[circle, draw=blue, fill=lightblue, minimum size=1cm] (rsi) at (8,0) {$\xy{RSI}$};
    \node[circle, draw=red, fill=lightred, minimum size=1cm] (bso) at (-4,-2) {$\xy{BSO}$};
    \node[circle, draw=red, fill=lightred, minimum size=1cm] (prv) at (-2,-2) {$\xy{PRV}$};
    \node[circle, draw=red, fill=lightred, minimum size=1cm] (hex4) at (0,-2) {$\xy{HEX}$};
    \node[circle, draw=red, fill=lightred, minimum size=1cm] (hlr) at (2,-2) {$\xy{HLR}$};
    \node[circle, draw=red, fill=lightred, minimum size=1cm] (hex5) at (4,-2) {$\xy{HEX}$};
    \node[circle, draw=red, fill=lightred, minimum size=1cm] (tur) at (6,-2) {$\xy{TUR}$};
    \node[circle, draw=red, fill=lightred, minimum size=1cm] (wsp) at (8,-2) {$\xy{WSP}$};
    \node[circle, draw=red, fill=lightred, minimum size=1cm] (hlf) at (10,-2) {$\xy{HLF}$};
    \node[circle, draw=red, fill=lightred, minimum size=1cm] (hex6) at (12,-2) {$\xy{HEX}$};
    \node[circle, draw=red, fill=lightred, minimum size=1cm] (acp) at (14,-2) {$\xy{ACP}$};
    \node[circle, draw=red, fill=lightred, minimum size=1cm] (bsi) at (16,-2) {$\xy{BSI}$};

    \draw (hex1) -- (hex4);
    \draw (hex2) -- (hex5);
    \draw (hex3) -- (hex6);
    \draw[arrow] (rso) -- (hex1);
    \draw[arrow] (hex1) -- (hex2);
    \draw[arrow] (hex2) -- (hex3);
    \draw[arrow] (hex3) -- (rsi);
    \draw[arrow] (bso) -- (prv);
    \draw[arrow] (prv) -- (hex4);
    \draw[arrow] (hex4) -- (hlr);
    \draw[arrow] (hlr) -- (hex5);
    \draw[arrow] (hex5) -- (tur);
    \draw[arrow] (tur) -- (wsp);
    \draw[arrow] (wsp) -- (hlf);
    \draw[arrow] (hlf) -- (hex6);
    \draw[arrow] (hex6) -- (acp);
    \draw[arrow] (acp) -- (bsi);
    
    \end{tikzpicture}

%% file: input/results/tms5.tex
\begin{tikzpicture}[scale=\tmsscale, transform shape]
    \node[circle, draw=red, fill=lightred, minimum size=1cm] (bso) at (-2,0) {$\xy{BSO}$};
    \node[circle, draw=red, fill=lightred, minimum size=1cm] (prv) at (0,0) {$\xy{PRV}$};
    \node[circle, draw=red, fill=lightred, minimum size=1cm] (acp) at (2,0) {$\xy{ACP}$};
    \node[circle, draw=red, fill=lightred, minimum size=1cm] (hex1) at (4,0) {$\xy{HEX}$};
    \node[circle, draw=red, fill=lightred, minimum size=1cm] (hex2) at (6,0) {$\xy{HEX}$};
    \node[circle, draw=red, fill=lightred, minimum size=1cm] (tur) at (8,0) {$\xy{TUR}$};
    \node[circle, draw=red, fill=lightred, minimum size=1cm] (wsp) at (10,0) {$\xy{WSP}$};
    \node[circle, draw=red, fill=lightred, minimum size=1cm] (hlr) at (12,0) {$\xy{HLR}$};
    \node[circle, draw=red, fill=lightred, minimum size=1cm] (hlf) at (14,0) {$\xy{HLF}$};
    \node[circle, draw=red, fill=lightred, minimum size=1cm] (bsi) at (16,0) {$\xy{BSI}$};
    \node[circle, draw=blue, fill=lightblue, minimum size=1cm] (rsi) at (3,-2) {$\xy{RSI}$};
    \node[circle, draw=blue, fill=lightblue, minimum size=1cm] (hex3) at (5,-2) {$\xy{HEX}$};
    \node[circle, draw=blue, fill=lightblue, minimum size=1cm] (hex4) at (7,-2) {$\xy{HEX}$};
    \node[circle, draw=blue, fill=lightblue, minimum size=1cm] (rso) at (9,-2) {$\xy{RSO}$};

    \draw (hex1) -- (hex3);
    \draw (hex2) -- (hex4);
    \draw[arrow] (bso) -- (prv);
    \draw[arrow] (prv) -- (acp);
    \draw[arrow] (acp) -- (hex1);
    \draw[arrow] (hex1) -- (hex2);
    \draw[arrow] (hex2) -- (tur);
    \draw[arrow] (tur) -- (wsp);
    \draw[arrow] (wsp) -- (hlr);
    \draw[arrow] (hlr) -- (hlf);
    \draw[arrow] (hlf) -- (bsi);
    \draw[arrow] (rso) -- (hex4);
    \draw[arrow] (hex4) -- (hex3);
    \draw[arrow] (hex3) -- (rsi);
\end{tikzpicture}

%% file: input/results/tms6.tex
\begin{tikzpicture}[scale=\tmsscale, transform shape]
    \node[circle, draw=red, fill=lightred, minimum size=1cm] (bso) at (-2,0) {$\xy{BSO}$};
    \node[circle, draw=red, fill=lightred, minimum size=1cm] (prv) at (0,0) {$\xy{PRV}$};
    \node[circle, draw=red, fill=lightred, minimum size=1cm] (hex1) at (2,0) {$\xy{HEX}$};
    \node[circle, draw=red, fill=lightred, minimum size=1cm] (acp) at (4,0) {$\xy{ACP}$};
    \node[circle, draw=red, fill=lightred, minimum size=1cm] (hlf) at (6,0) {$\xy{HLF}$};
    \node[circle, draw=red, fill=lightred, minimum size=1cm] (hex2) at (8,0) {$\xy{HEX}$};
    \node[circle, draw=red, fill=lightred, minimum size=1cm] (tur) at (10,0) {$\xy{TUR}$};
    \node[circle, draw=red, fill=lightred, minimum size=1cm] (wsp) at (12,0) {$\xy{WSP}$};
    \node[circle, draw=red, fill=lightred, minimum size=1cm] (bsi) at (14,0) {$\xy{BSI}$};
    \node[circle, draw=blue, fill=lightblue, minimum size=1cm] (rso) at (1,-2) {$\xy{RSO}$};
    \node[circle, draw=blue, fill=lightblue, minimum size=1cm] (hex3) at (3,-2) {$\xy{HEX}$};
    \node[circle, draw=blue, fill=lightblue, minimum size=1cm] (hlr) at (5,-2) {$\xy{HLR}$};
    \node[circle, draw=blue, fill=lightblue, minimum size=1cm] (hex4) at (7,-2) {$\xy{HEX}$};
    \node[circle, draw=blue, fill=lightblue, minimum size=1cm] (rsi) at (9,-2) {$\xy{RSI}$};

    \draw (hex1) -- (hex3);
    \draw (hex2) -- (hex4);
    \draw[arrow] (bso) -- (prv);
    \draw[arrow] (prv) -- (hex1);
    \draw[arrow] (hex1) -- (acp);
    \draw[arrow] (acp) -- (hlf);
    \draw[arrow] (hlf) -- (hex2);
    \draw[arrow] (hex2) -- (tur);
    \draw[arrow] (tur) -- (wsp);
    \draw[arrow] (wsp) -- (bsi);
    \draw[arrow] (rso) -- (hex3);
    \draw[arrow] (hex3) -- (hlr);
    \draw[arrow] (hlr) -- (hex4);
    \draw[arrow] (hex4) -- (rsi);
\end{tikzpicture}

%% file: input/results/exp1.tex
\subsubsection{Understanding Model Performance Through Known Set Size and Training Epochs}

The initial step focuses on identifying the desired size of the $\mathcal{G}_{known}$ and the appropriate number of epochs for training our model. 
Given the need to balance model efficacy and data collection costs, establishing these parameters at the outset is essential.
This first experiment explores models using five different $N_{known}$, each with seven independent runs for statistically significant results.
The optimal dataset size and number of training epochs can be established by gathering and analyzing the data from these iterations and training runs. 
This determination will be based on charting the average training loss observed in each iteration. 
Additionally, the decision-making process will consider other vital metrics, such as the mean F1 score and the aggregate accuracy when the model is applied to the unseen data.

\begin{table}[p]
\centering
\caption{The model metrics for different $N_{known}$ averaged over the seven independent runs.}
\begin{tabular}{rccccc}
\hline \hline
Known Size \% & Mean Accuracy & Mean Precision & Mean Recall & Mean F1 & Mean AUC\\
\hline
20\phantom{.00} & 0.915 & 0.947 & 0.854 & 0.864 & 0.860\\
10\phantom{.00} & 0.909 & 0.954 & 0.838 & 0.853 & 0.845\\
5\phantom{.00} & 0.902 & 0.940 & 0.813 & 0.840 & 0.845\\
2.5\phantom{0} & 0.881 & 0.930 & 0.760 & 0.790 & 0.825\\
1.25 & 0.866 & 0.915 & 0.703 & 0.744 & 0.824\\
\hline \hline
\end{tabular}
\label{tab:tab2}
\end{table}

\begin{table}[p]
    \centering
    \caption{Precision, Recall, and F1 scores broken down by label.}
    \resizebox{\textwidth}{!}{\begin{tabular}{rcccc}
        \hline \hline
        & Will not compile ($l_1$) & Will compile ($l_2$) & Will not simulate ($l_3$) & Will simulate ($l_4$)\\
        \hline
        \begin{tabular}{r}
           Known Size \% \\ \hline 20\phantom{.00} \\ 10\phantom{.00}\\ 5\phantom{.00}\\ 2.5\phantom{0} \\ 1.25 
        \end{tabular}
        &
        \begin{tabular}{ccc} 
            Precision & Recall & F1\\
            \hline
            0.948 & 0.965 & 0.957\\
            0.941 & 0.968 & 0.954\\
            0.937 & 0.964 & 0.951\\
            0.941 & 0.948 & 0.945\\
            0.883 & 0.982 & 0.930\\
        \end{tabular} &
        \begin{tabular}{ccc}
            Precision & Recall & F1\\
            \hline
            0.818 & 0.743 & 0.778\\
            0.819 & 0.708 & 0.759\\
            0.800 & 0.688 & 0.740\\
            0.740 & 0.714 & 0.727\\
            0.810 & 0.373 & 0.510\\
            
        \end{tabular} &
        \begin{tabular}{ccc}
            Precision & Recall & F1\\
            \hline
            0.981 & 0.984& 0.982\\
            0.979 & 0.983& 0.981\\
            0.974 & 0.988& 0.981\\
            0.959 & 0.992& 0.975\\
            0.964 & 0.986& 0.975\\            
        \end{tabular} &
        \begin{tabular}{ccc}
            Precision & Recall & F1\\
            \hline
            0.758 & 0.722& 0.739\\
            0.744 & 0.692& 0.717\\
            0.784 & 0.610& 0.686\\
            0.769 & 0.385& 0.512\\
            0.702 & 0.471& 0.562\\            
        \end{tabular}\\
        \hline\hline
    \end{tabular}}
    \label{tab:tab3}
\end{table}

\begin{table}[p]
    \centering
    \caption{Matthews Correlation Coefficient for each $N_{known}$ and label pairing.}
    \begin{tabular}{rccc}
        \hline\hline
        Known Size \% & $l_1$ \& $l_2$ & $l_3$ \& $l_4$\\
        \hline
        20\phantom{.00} & 0.736 & 0.722\\
        10\phantom{.00} & 0.717 & 0.699\\
        5\phantom{.00} & 0.693 & 0.673\\
        2.5\phantom{0} & 0.671 & 0.524\\
        1.25 & 0.496 & 0.552\\
        \hline\hline
    \end{tabular}
    \label{tab:mcc}
\end{table}

\begin{figure}[p]
    \centering
    \includegraphics[scale=0.75]{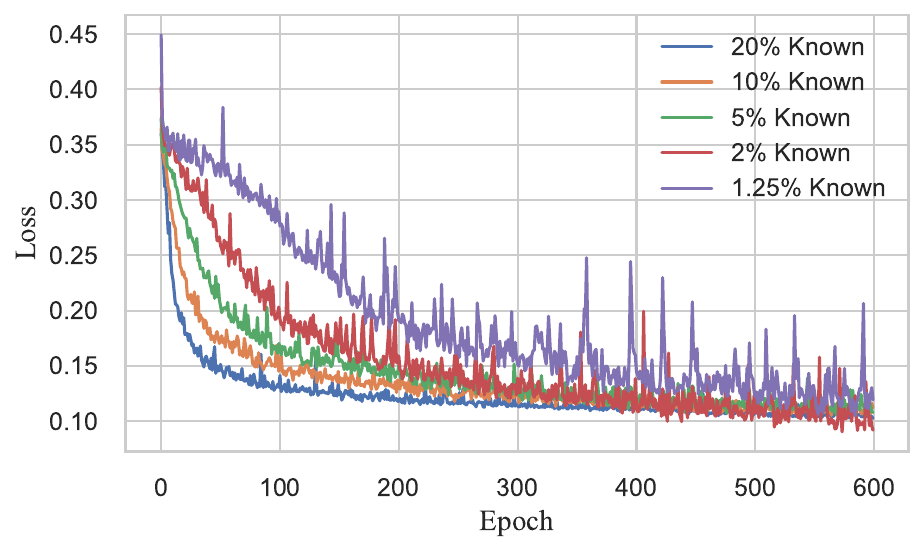}
    \caption{Average training loss for each iteration per run.}
    \label{fig:tloss}
\end{figure}

The metrics for each iteration and run of the model are detailed in Table~\ref{tab:tab2}.
It is evident from these results that we can produce a performant model even when reduced to smaller values of around 5\% of $N_{all}$. 
Upon reviewing \tref{tab:tab3}, we find a detailed analysis of precision, recall, and F1 scores segmented by label. 
It becomes apparent that label \(l_1\) maintains robust performance throughout all the explored $N_{known}$. 
This strong performance may be due to a large portion of the data being labeled 1 within this category. 
Additionally, when there's a considerable imbalance in the data for specific labels, such as \(l_2\) and \(l_4\), the model can exhibit issues with very small $N_{known}$.
Finally, by examining the Matthews Correlation Coefficient (MCC) for each label pairing as presented in \tref{tab:mcc}, the model exhibits weaker performance with smaller known dataset sizes, consistent with expectations.

Now, by examining \fref{fig:tloss}, which illustrates the average training loss for each iteration across different runs, we observe that the loss remains relatively consistent between 20\% and 2.5\% dataset sizes.
Furthermore, the training loss seems to slow down around 300 epochs in most cases.

Integrating these insights enables an informed selection of the $N_{known}$ and epoch count for subsequent experiments.
Based on this data, it was decided to utilize 5\% of the dataset and conduct training over 300 epochs. 
This decision is supported by the data in \tref{tab:tab2}, which indicates that the average accuracy through to the AUC remains high for this dataset size. 
Furthermore, as detailed in \tref{tab:tab3}, the 5\% dataset size demonstrates strong performance even when analyzed by an individual label.
Lastly, the 5\% known set size MCC value shows relatively high prediction strength for all labels.
The choice of 300 epochs is justified by observing that the loss for each model tends to level off around this point, with the differences between them being negligible.

\subsubsection{Improving Model Performance with Feature Engineering}
\label{sec:featureeng}
In this section, we aim to enrich the feature set \(\mathbf{X}\) by adding graph-based metrics beyond the conventional vertex labels (e.g., $\xy{HEX}$, $\xy{BSO}$, $\xy{BSI}$). 
Integrating these additional features, a process known as feature engineering and feature selection aims to improve the model's ability to capture variance within the training dataset. 
We hypothesize that incorporating these new attributes may enhance the model's performance without significantly increasing the number of training epochs required. 
However, a potential challenge arises when the structural characteristics of various graph candidates are closely similar, leading to analogous graph-based features. 
To address this, \textit{Principal Component Analysis (PCA)} is employed as a solution.

\begin{figure}[h]
    \centering
    \begin{subfigure}[t]{0.45\textwidth}
        \includegraphics[scale=0.45]{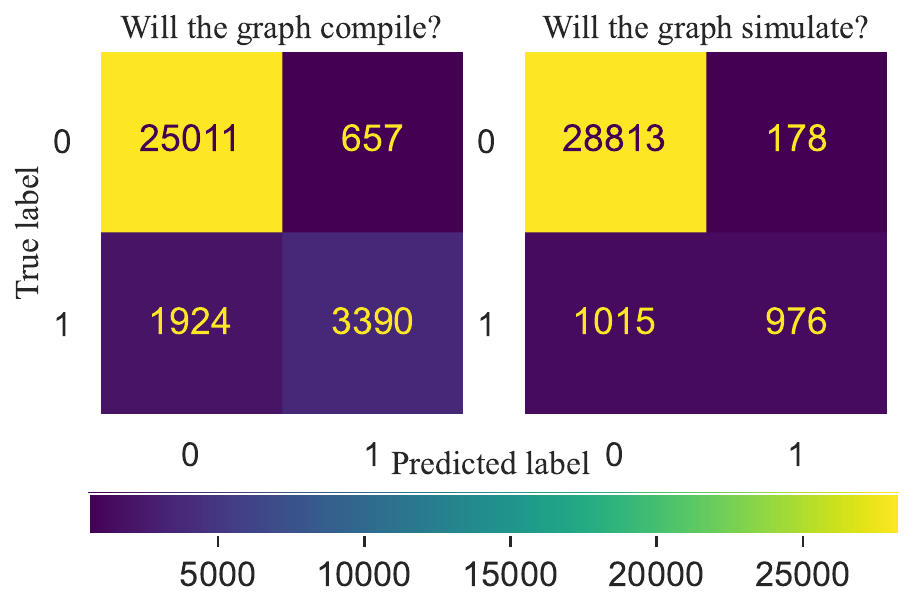}
        \caption{Baseline confusion matrix.}
        \label{fig:baselinecm}
    \end{subfigure}
    \hfill
    \begin{subfigure}[t]{0.45\textwidth}
        \includegraphics[scale=0.45]{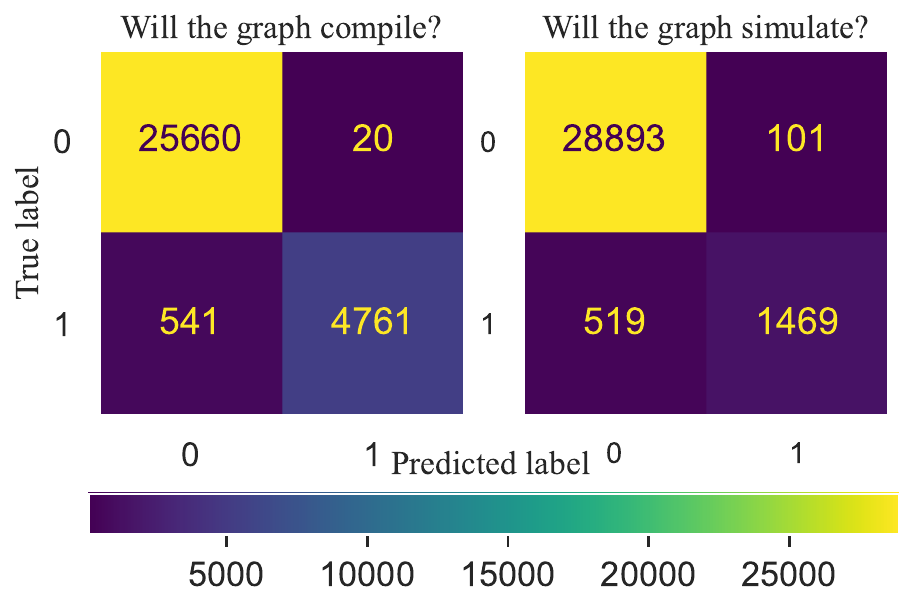}
        \caption{With the addition of PCA features.}
        \label{fig:harmonic_cm}
    \end{subfigure}
    \caption{Confusion matrices on $\mathcal{G}_{unknown}$ for models with different feature sets.}
\end{figure}

We note that the computational cost of incorporating these extra features is relatively minimal compared to the calculation of $J(G_i)$, meaning they only slightly increase the overall processing time for each graph. 
With the addition of these new features, the feature matrix $\textbf{X}$ for each graph expands from a dimensionality of $\mathbb{R}^{n \times 1}$ to $\mathbb{R}^{n \times 2}$. 
The base model, without adding the extra feature, was run once more at 5\% the known data for 300 epochs, and the confusion matrix is shown in \fref{fig:baselinecm}.
Using the values from the confusion matrix, the model achieved an accuracy of 91.67\%. 
The MCC values are 68.5\% and 62.7\%, respectively.
The confusion matrix for the model with the additional feature, detailed in \fref{fig:harmonic_cm}, reveals significant improvements: the model's accuracy reached 98.19\% and 98.0\%, and the MCC 93.54\% and 82.17\%. These enhancements indicate that the model achieved superior predictive performance on the same $\mathcal{G}_{unknown}$ dataset, with all metrics showing better results. 
Consequently, we will incorporate this feature in future model constructions.

\begin{figure}[t]
    \centering
    \begin{subfigure}[t]{0.45\textwidth}
        \includegraphics[scale=0.45]{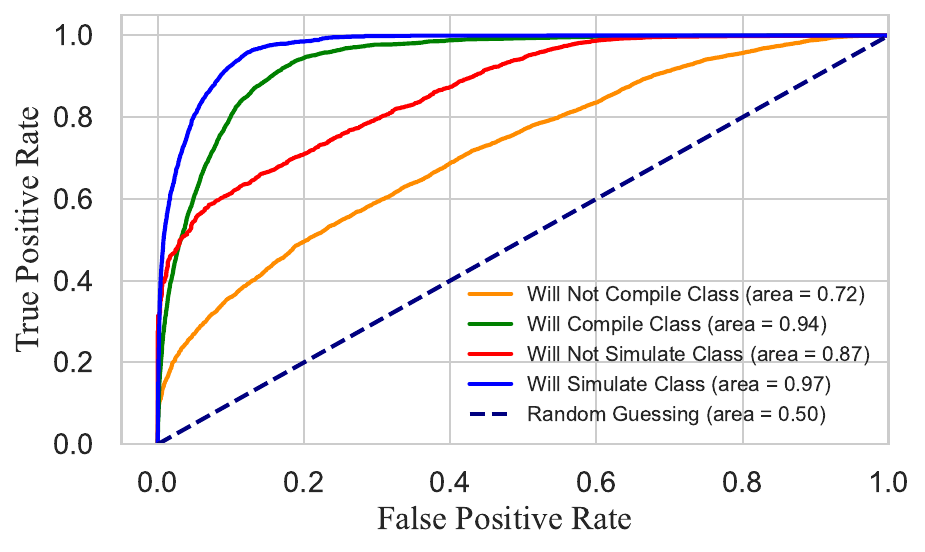}
        \caption{Baseline ROC curve.}
    \end{subfigure}
    \hfill
    \begin{subfigure}[t]{0.45\textwidth}
        \includegraphics[scale=0.45]{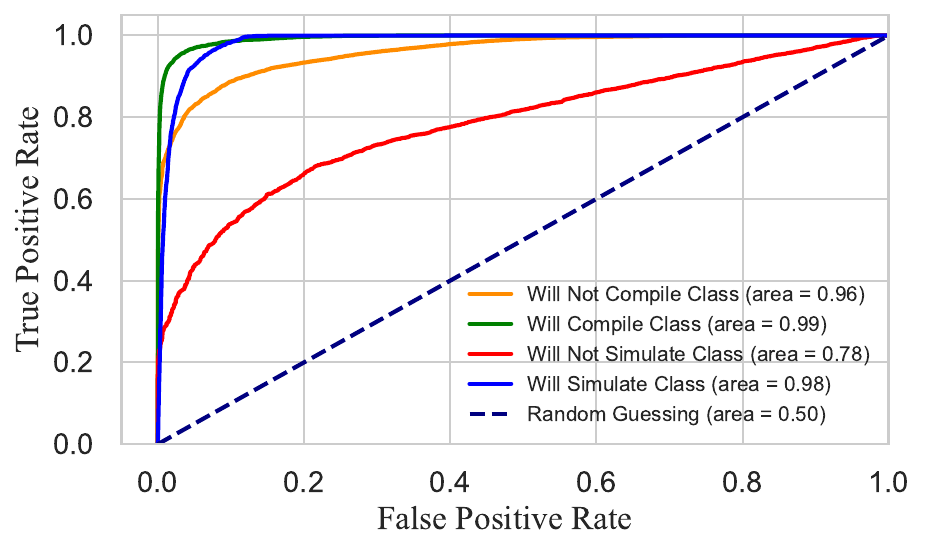}
        \caption{With the addition of Harmonic Centrality.}
    \end{subfigure}
    \caption{ROC curves on $\mathcal{G}_{unknown}$ for models with different feature sets.}
    \end{figure}

%% file: input/results/exp2.tex
\subsection{Iterative Down-selection Based on Graph Performance}
\label{sec:iterclass}


In addition to compiling and simulating the TMS graphs, we are also concerned with which ones perform the best concerning Eq.~(\ref{eq:J}).
In this section, we use the iterative classification approach from \sref{sec:iterative-class} to down-select a subset of graphs from $\mathcal{G}_{all}$ with higher median performance (i.e., potentially good graphs).



\begin{figure}[t]
    \centering
    \includegraphics[scale=0.8]{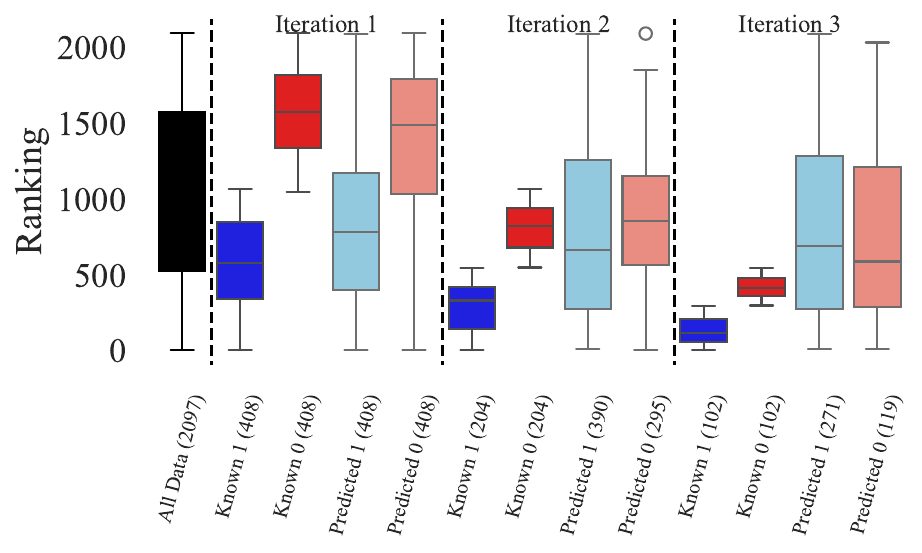}
    \caption{Three cycles of the method outlined in \sref{sec:iterative-class} are applied to retrain a GDL model, each cycle using progressively refined subsets of graphs to minimize performance issues. These subsets are indicated by their respective graph counts. Additionally, all data point rankings are distributed for comparative analysis.}
    \label{fig:iterclass}
\end{figure}

Here, we start with 40\% of the graphs already evaluated as presented in \fref{fig:iterclass}. 
We see the ranking of values from each set from each iteration compared to the rankings from the original dataset $\mathcal{G}_{all}$.
Initially, the far-left bar plot illustrates the distribution of $J(\mathcal{G}_{all})$, ranking its values from 1 (the best $J$ value) to $N_{all}$.
The subsequent two bar plots demonstrate the division of $\mathcal{G}_{known}$ according to the median performance classification threshold in the first iteration. 
This iteration's last two bar plots represent the GDL model's predictions of 1s and 0s. 
It's important to note that some incorrectly classified graphs exist, yet the medians remain distinct and aligned with their respective ``Known'' categories. 
At the end of iteration 3, the final ``good'' set had 373 graphs, with a median of 282.01, much lower than the initial classification of 549.53.
For this final set produced in iteration 3, the model predicted 4 of the top 10, 69 of the top 100, and 133 of the top 200, with an overall count of 832 out of 2098 evaluated graphs.

An additional nine runs were conducted to gain a deeper insight into the statistical significance of this iterative method, each with a different random subset from $\mathcal{G}_{known}$. 
Across these runs, the average number of optimized graphs was 875.7, with a standard deviation of 34.3. 
Specifically, the average for the top 10 graphs was 4.75 with a standard deviation of 1.5; for the top 100, it was 69.0 with a standard deviation of 4.95.
Consequently, these results show fair performance when identifying these desired sets but does outperform previous attempts.

\begin{figure}[t]
    \centering
    \begin{subfigure}[t]{0.49\textwidth}
        \includegraphics[width=\textwidth]{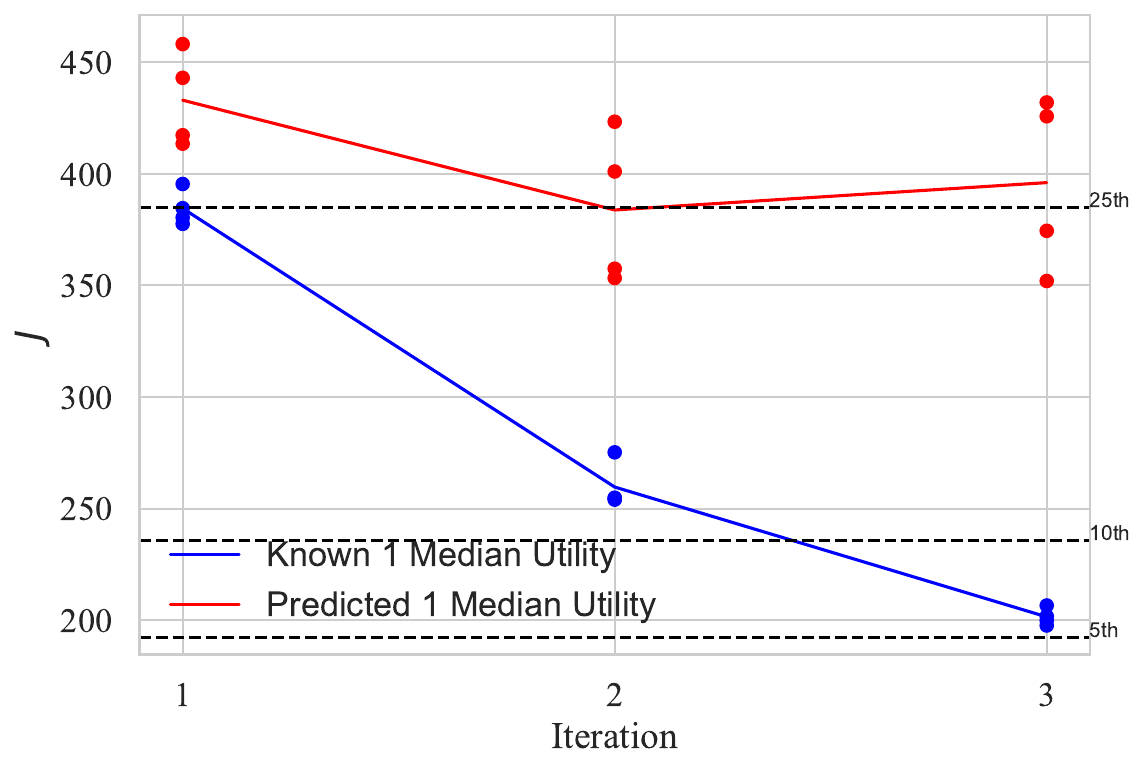}
        \caption{Median utility values of the ``Known 1'' and ``Predicted 1'' sets averaged over ten runs.}
        \label{fig:iter_avg}
    \end{subfigure}
    \hfill
    \begin{subfigure}[t]{0.49\textwidth}
        \includegraphics[width=\textwidth]{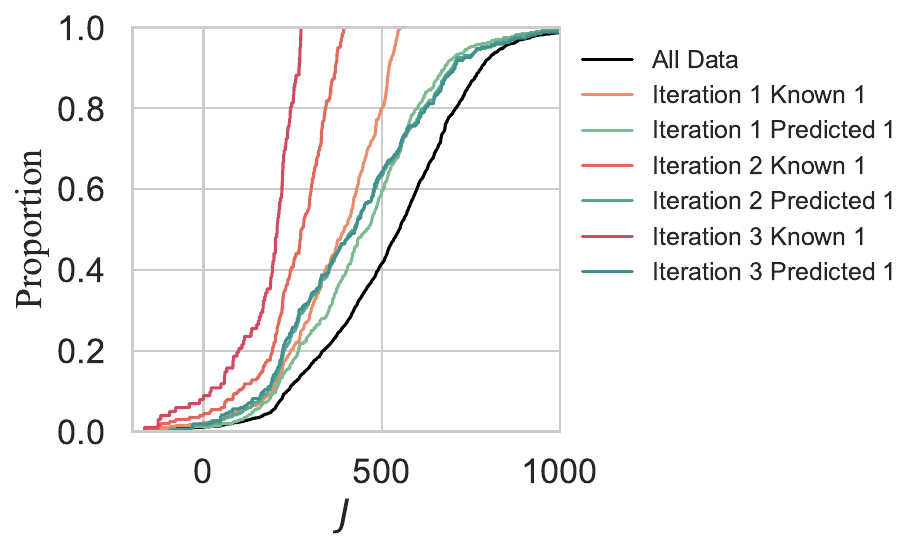}
        \caption{The cumulative distribution for a single run displays the proportion of values contained in each iteration.}
        \label{fig:ecdf}
    \end{subfigure}
    \caption{Some results from the iterative classification for down-selection study.}
\end{figure}

Finally, in \fref{fig:iter_avg}, we can see both the ``Known 1'' and ``Predicted 1'' categories averaged over the ten runs out to three iterations.
In the third iteration, the model achieved poor classification results exhibited by the significant variation in median utility values, primarily due to the very small training set size at this point.
Figure~\ref{fig:ecdf} displays the cumulative distribution of values across each iteration. 
As evident in conjunction with \fref{fig:iter_avg}, it can be observed that with every iteration, the values in both the ``Known 1'' and ``Predicted 1'' sets diminish, increasingly concentrating on the better-performing graphs.
In this case, the observed reduction in median values suggests that the \textit{iterative GDL} approach is effectively refining the selection of potentially ``good'' graphs, focusing on those that rank in the higher percentiles of \(J\).

%% file: input/conclusion.tex
This study utilizes the Iterative Classification of Graph-Set-Based Design framework (IC-GSBD) to classify and refine graph-based aircraft thermal management systems (TMSs), aiming to identify feasible and superior solutions in an engineering design context. 
In the case study, all graph objects were known, but certain outcomes, such as model simulatability and performance, were computationally expensive to determine. 
Thus, the effectiveness of IC-GSBD in distinguishing ``good'' and ``bad'' graphs with minimal known outcome data was demonstrated.
Moreover, incorporating the graph-based features of harmonic, betweenness, and eigenvector centrality, as well as spectral radius utilizing the principle component analysis (PCA) method, has enhanced several critical metrics.

When determining a graph's compilability and simulatability, models have achieved over 97\% accuracy while utilizing only 5\% of the entire dataset ($\mathcal{G}_{all}$).
For the iterative down-selection based on graph performance, using around 40\% of the data identified on average 7.3 (1.79) of the top 10 graphs and 75.5 (7.95) of the top 100.
With the extreme computational costs associated with determining any of these outcomes on the set of graphs, the results demonstrate a potentially drastic reduction in total costs.

Currently, the outcomes presented serve as a step towards in exploring general strategies for creating efficient GDL models for graph-based engineering design problems tailored for the classification and down-selection goals.
Critical aspects still require further study, especially when it comes to smaller datasets and intelligently selecting which graphs to evaluate outcomes (rather than the pure random approach used in this paper).
Future research directions include merging the two main GDL tasks for this same case study in a workflow that uses the first compilability/simulatability GDL model to select graphs for performance evaluation and down-selection.
There is also interest in exploring more regression-based approaches for predicting graph performance (and combining it with the quite accurate compilability/simulatability GDL classification model).
Another key area can involve the multi-objective nature of Eq.~(\ref{eq:J}). Rather than combining all of these metrics of interest into one, we might desire the Pareto set of solutions, which could include many instances of the same graph with different parameter values.
This could require innovatively adding other outcome information, such as simulated temperature values and mass flow rates, to the graph data object.